\documentclass[10pt,aps,prd,amsmath,showpacs,twocolumn,floatfix]{revtex4-1}
\usepackage{graphicx}
\usepackage{hyperref}
\usepackage{bm}
\usepackage{natbib}

\renewcommand{\vec}[1]{\bm{#1}}
\newcommand{\vcm}{\vec v_{\mathrm{cm}}}
\newcommand{\tint}{t_{\mathrm{int}}}

\begin{document}
\title{Shape of Cosmic String Loops}

\author{Craig J Copi}
\homepage{http://www.phys.cwru.edu/projects/strings/}
\affiliation{CERCA, Department of Physics, 
Case Western Reserve University, Cleveland, OH 44106-7079}
\author{Tanmay Vachaspati}
\affiliation{Physics Department, Arizona State University, Tempe, AZ 85287}

\begin{abstract}
  Complicated cosmic string loops will fragment until they reach simple,
  non-intersecting (``stable'') configurations.  Through extensive
  numerical study we characterize these attractor loop shapes including 
  their length, velocity, kink, and cusp distributions. We find that 
  an initial loop
  containing $M$ harmonic modes will, on average, split into $3M$ stable
  loops.  These stable loops are approximately described by the degenerate
  kinky loop, which is planar and rectangular, independently of the number of 
  modes on the initial loop.  This is confirmed by an analytic construction 
  of a stable family of perturbed degenerate kinky loops. The average stable 
  loop is also found to have a $40\%$ chance of containing a cusp.
  We examine the properties of stable loops of different lengths and find
  only slight variation.  
  Finally we develop a new analytic scheme to explicitly solve the string
  constraint equations.
\end{abstract}
\pacs{98.80.Cq, 11.27.+d}

\maketitle

\section{Introduction}
\label{sec:introduction}

Cosmic strings may be observed through their gravitational wave emission,
gravitational lensing of background galaxies, imprints on the cosmic
microwave background, particle emission, and various other signatures. To
work out the signatures of cosmic strings it is essential to know the
properties of the string network, the number density of loops and their
length and shape distributions. Several observational signatures, including
all burst events, whether gravitational or other, depend on the sudden
whip-like motion of the string, a feature called a ``cusp''. Yet other
signatures are affected by sharp corners on a string, called ``kinks''.
(See Ref.~\cite{Vilenkin-Shellard1994} for a review.)  Hence, to make
reliable observational predictions from cosmic strings, it is important to
characterize the generic properties of cosmic string loops and to quantify
the frequency with which kinks and cusps occur.

Scherrer and Press (SP) \cite{Scherrer1989} undertook a study of the
dynamics of large loops starting with randomly chosen loops containing 10
harmonic modes. As each loop oscillates, it self-intersects and reconnects,
breaking into smaller loops. In SP the loop is approximated by 128 discrete
segments thus limiting the resolution of their study. The study in SP was
followed up by Scherrer, Quashnock, Spergel and Press (SQSP)
\cite{Scherrer1990} with a focus on evaluating the gravitational power
emitted from string loops.  A similar study was undertaken by Casper and
Allen (CA) \cite{Casper1995} and is the most extensive study done to
date. CA considered loops with 10 harmonics and discretized the loop using
600 segments.

In the present work we have built upon the SP scheme, improvements coming
from more computational power and choice of algorithms. To put our current
study in perspective, we have considered loops with 3 to 50 harmonics and
discretized using several times 10,000 segments. Our very fine resolution
ensures that discretization effects are unimportant as will be clear from
our results.  A comparison of previous loop fragmentation studies with this
one is given in Table~\ref{tab:loop-parameters}.

\begin{table}[b]
  \begin{ruledtabular}
    \begin{tabular}{crcrr} \\
      & & \textbf{Modes} & \multicolumn{1}{c}{\textbf{Initial}}
      & \multicolumn{1}{c}{\textbf{Stable}} \\
      \multicolumn{1}{c}{\textbf{Source}} &
      \multicolumn{1}{c}{\textbf{Segments}} & \multicolumn{1}{c}{($M$)} &
      \multicolumn{1}{c}{\textbf{Loops} ($N$)} &
      \multicolumn{1}{c}{\textbf{Loops}}  \\ \hline
      SP & 128\footnote{Loops were also run with 256 segments without
        producing significantly more small loops} & 10 & 20 & 561 \\
      CA & 600\footnote{The loops were rerun with 800 segments resulting in
        almost no new daughter loops being produced.} & 10 & 200 & 5,723 \\
      Present work & 10,000 & 3 & 1,000 & 8,308 \\
       --- & 10,000 & 10 & 3,000 & 94,628 \\
       --- & 10,000 & 20 & 1,000 & 63,490 \\
       --- & 10,000 & 30 & 1,000 & 96,207 \\
       --- & 10,000 & 40 & 1,000 & 128,764 \\
       --- & 10,000 & 50 & 1,000 & 157,968 \\
       --- & 50,000 & 10 & 1,000 & 32,158 \\
       --- & 50,000 & 50 & 1,000 & 162,157 \\
    \end{tabular}
  \end{ruledtabular}
  \caption{Parameters from studies of loop fragmentation.  Modes refers to
    $M$ in Eq.~(\ref{eqn:loopexpansion}). The numbers provided here are for
    the ``Type A'' loops defined by SP and used by CA; see the discussion
    after Eq.~(\ref{eqn:loopexpansion}) for more details.  For SP see
    \cite{Scherrer1989} and for CA see \cite{Casper1995}.}
  \label{tab:loop-parameters}
\end{table}

Our conclusions can be summarized as follows. Large loops with many
oscillation harmonics split into smaller loops until the small loops have a
\textit{minimal} number of harmonics. In other words, the number of
fragments is directly proportional to the number of harmonics on the
initial loop; on average $3M$ loops are created from an initial loop with
$M$ modes. The final non-self-intersecting (``stable'') loops are
approximately planar and their left- and right- moving modes tend to be
orthogonal. The average stable loop contains four kinks and has a $40\%$
chance of containing a cusp. Hence, realistic cosmic string loops should be
visualized as oscillating rectangular shapes, like the ``degenerate kinky
loops'' discussed in Ref.~\cite{Garfinkle1987}. This picture is
strikingly clear in the animations available in Ref.~\cite{copi-strings}.

The conclusion that stable loops are close to degenerate kinky loops
also means that gravitational and other radiation from stable loops 
can be estimated by the radiation from degenerate kinky loops, which
is simple to calculate analytically \cite{Garfinkle1987}. Knowing the
shape of loops can also help with other observational signatures
{\it e.g.} lensing.

Our detailed results can be found in Sec.~\ref{sec:results}.  At a finer
level, the shape of a cosmic string loop may also depend on its length. We
investigate this question in Sec.~\ref{sec:lwa} and find only weak
dependence. It is a fair approximation to think of long loops as also being
approximately planar and rectangular.  In Sec.~\ref{sec:discussion} we give
analytical arguments that help us understand our numerical results. Notable
is our demonstration of a set of perturbed degenerate kinky loops that are
stable. We conjecture that such loops are attractors for the evolution of
cosmic string loops.

In the Appendices we describe technical details of the numerical algorithm,
formulas to boost to the rest frame of a loop, and finally, a new scheme to
explicitly solve the Nambu-Goto equations and the string constraint
equations.  This new scheme is very suitable for constructing cosmic string
loops but in our numerical work we use the SP scheme to aid comparison to
previous work.

\section{Procedure}
\label{sec:procedure}

A string in flat spacetime is described by $X^\mu = (t,{\vec x}(\sigma,t))$
where $\sigma$ is the parameter along the string and $t$ is Minkowski time.
The solution of the Nambu-Goto string equations yields independent left-
and right-moving modes
\begin{equation}
  {\vec x}(\sigma , t) = \frac{1}{2} [ {\vec a}(\sigma_-) +
    {\vec b}(\sigma_+) ],
  \label{eqn:nambugoto}
\end{equation}
where $\sigma_\pm = \sigma\pm t$. We choose equal intervals of $\sigma$ to
label equal amounts of energy which gives the additional constraint
\begin{equation}
  |{\vec p}| = 1 = |{\vec q}|
  \label{eqn:constraints}
\end{equation}
where ${\vec p} \equiv {\vec a}'$ and ${\vec q} \equiv {\vec b}'$.  Closure
of a loop of length $L$ implies that
\begin{equation}
  \int_0^{L} \vec p\,d\sigma  = - \int_0^{L} \vec q\,d\sigma 
  \label{eqn:closure}
\end{equation}
while, in the center of momentum of the loop, each of the two integrals
vanishes.  A helpful geometrical picture is that the vector functions
${\vec p}$ and ${\vec q}$ trace out a path on the surface of a two
dimensional sphere, also called the ``Kibble-Turok (KT) sphere''.

The velocity of a point on the string is $({\vec q}-{\vec p})/2$. Hence, at
points where the curves corresponding to ${\vec p}$ and $-{\vec q}$
intersect, the velocity of the string reaches the speed of light. Such
momentarily light-like points on the string are known as ``cusps''. Finally
note that ${\vec p}$ and ${\vec q}$ need not be continuous. If there is a
break in either of the curves, it implies a discontinuity in the tangent
vector to the string, which will appear as a sharp corner or ``kink''.

\begin{figure}
  \includegraphics[height=0.45\textwidth,angle=-90]{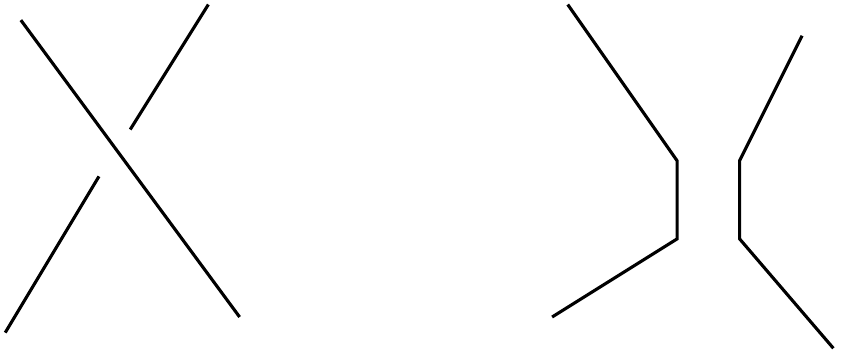}
  \caption{When two strings intersect, they reconnect in an
    ``intercommutation event with the production of four kinks
    (discontinuities in the tangent to the string), two on each string.  }
  \label{fig:intercommuting}
\end{figure}

The Nambu-Goto description of a cosmic string breaks down when two string
segments intersect. The Nambu-Goto dynamics has to be supplemented by the
condition that the strings intercommute, \textit{i.e.}\ reconnect, at the
point of intersection. This is sketched in Fig.~\ref{fig:intercommuting},
which also shows the four kinks that are created during intercommutation, two
on each string.

To numerically model strings we follow SP and choose initial loops from an
ensemble constructed by expanding the left- and right-movers in Fourier
series
\begin{eqnarray}
  a_i (s) &=& \sum_{m=1}^M a_{m,i} \cos (ms+\phi_{a,m,i}), 
  \nonumber \\
  b_i (s) &=& \sum_{m=1}^M b_{m,i} \cos (ms+\phi_{b,m,i}),
  \label{eqn:loopexpansion}
\end{eqnarray}
where $i$ stands for the three components of the vectors.  Here we have
taken $s\in[0,2\pi]$. The mode coefficient vectors, ${\vec a}_m$ and ${\vec
  b}_m$, are chosen so that each component of these vectors is uniformly
distributed in the interval $[0,1]$. The phases, $\phi_{a,m,i}$ and
$\phi_{b,m,i}$, are chosen to be uniformly distributed in the interval
$[0,2\pi]$. This corresponds to the ``type A'' scheme in SP and CA. Had we
chosen the length of the mode coefficient vectors in the interval
$[0,1/m^2]$, we would have a direct correspondence with the ``type B''
scheme of SP and CA. As already discussed in CA, the type A loops have more
power on small scales and leads to more fragmentations. For this reason we
have only studied type A loops in this work.  However, the shapes of the
final stable loop population is expected to be the same in both schemes. We
have checked this by considering several different values of $M$ since this
controls the power on small scales.  A comparison of our choice of
parameters with those of earlier work is given in
Table~\ref{tab:loop-parameters}.

The challenge now is to meet the constraints in
Eq.~(\ref{eqn:constraints}).  This is done by recognizing that we are free
to choose the parameter $s$ as a function of the parameter $\sigma$. By
differentiating Eq.~(\ref{eqn:loopexpansion}) we obtain
\begin{equation}
  1= |{\vec p}| = \left | \frac{ds}{d\sigma} \right | 
  \left | \sum_{m=1}^M {\vec a}_m m \sin (ms+\phi_{a,m}) \right |.
\label{eqn:diffeq}
\end{equation}
This provides a differential equation for $\sigma$ in terms of $s$ which we
solve numerically with the additional requirement that $ds/d\sigma \ge 0$.
To insure that $s(\sigma)\in[0,2\pi]$ for $\sigma\in[0,1]$ we first rescale
$\vec a_m$ and $\vec b_m$ before solving~(\ref{eqn:diffeq}).  A numerical
inversion then gives $s$ as a function of $\sigma$ which finally allows us
to obtain ${\vec p}(\sigma)$ and ${\vec q}(\sigma)$ at the initial time,
$t=0$. Note that the value of $M$ corresponds to the ``number of
harmonics'' on the loop in terms of the parameter $s$ but not in terms of
the invariant length $\sigma$. However we will still refer to $M$ as the
number of harmonics.

Once we know the function ${\vec p}(\sigma )$ and ${\vec q}(\sigma)$, the
Nambu-Goto evolution is straight-forward since only the arguments of the
functions change as given in Eq.~(\ref{eqn:nambugoto}).  The numerical
implementation of intercommutation events is more involved and described in
Appendix~\ref{app:loop-intercommutation}.

An intercommutation causes the original loop to fragment into two smaller
loops, with each of the two loops having one left-moving kink and one
right-moving kink. In this way, the loops keep fragmenting with the
production of more and more kinks. The fragmentation ceases when a loop
reaches a non-self-intersecting configuration. It is these stable loops
that we wish to study and characterize.

In principle, two fragments of the initial loop can intersect and reconnect
to form a bigger loop. In practice, since the fragments are flying apart,
such collisions do not happen frequently. In any case, we ignore such
mergers.

\section{Numerical Results}
\label{sec:results}

\begin{figure}
  \includegraphics[width=0.45\textwidth]{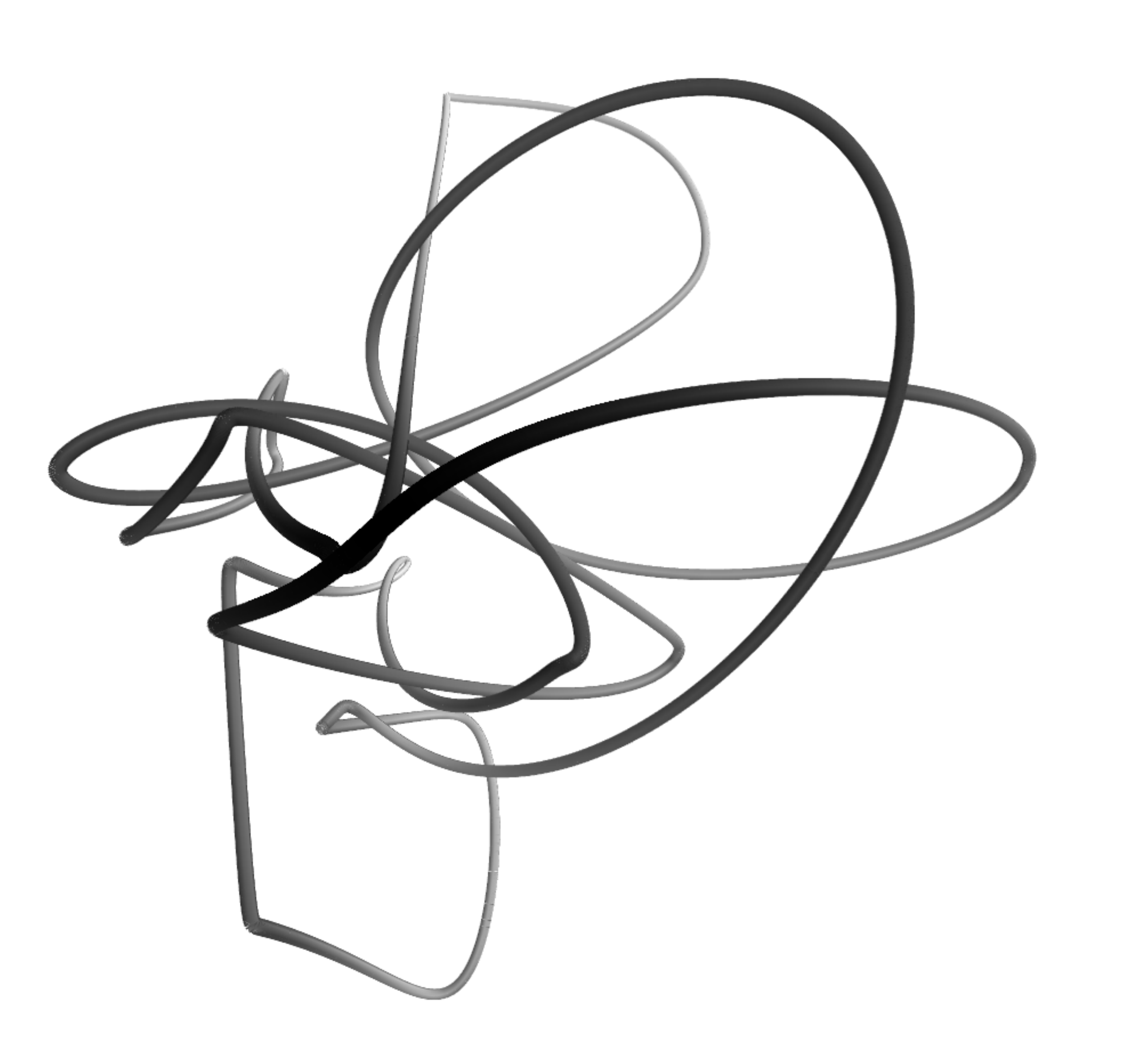}
  \caption{ A sample initial loop with 10 harmonic modes.  The darker,
    thicker segments of the loop are nearer the viewer.  }
  \label{fig:sampleloop}
\end{figure}

\begin{figure}
  \includegraphics[width=0.45\textwidth]{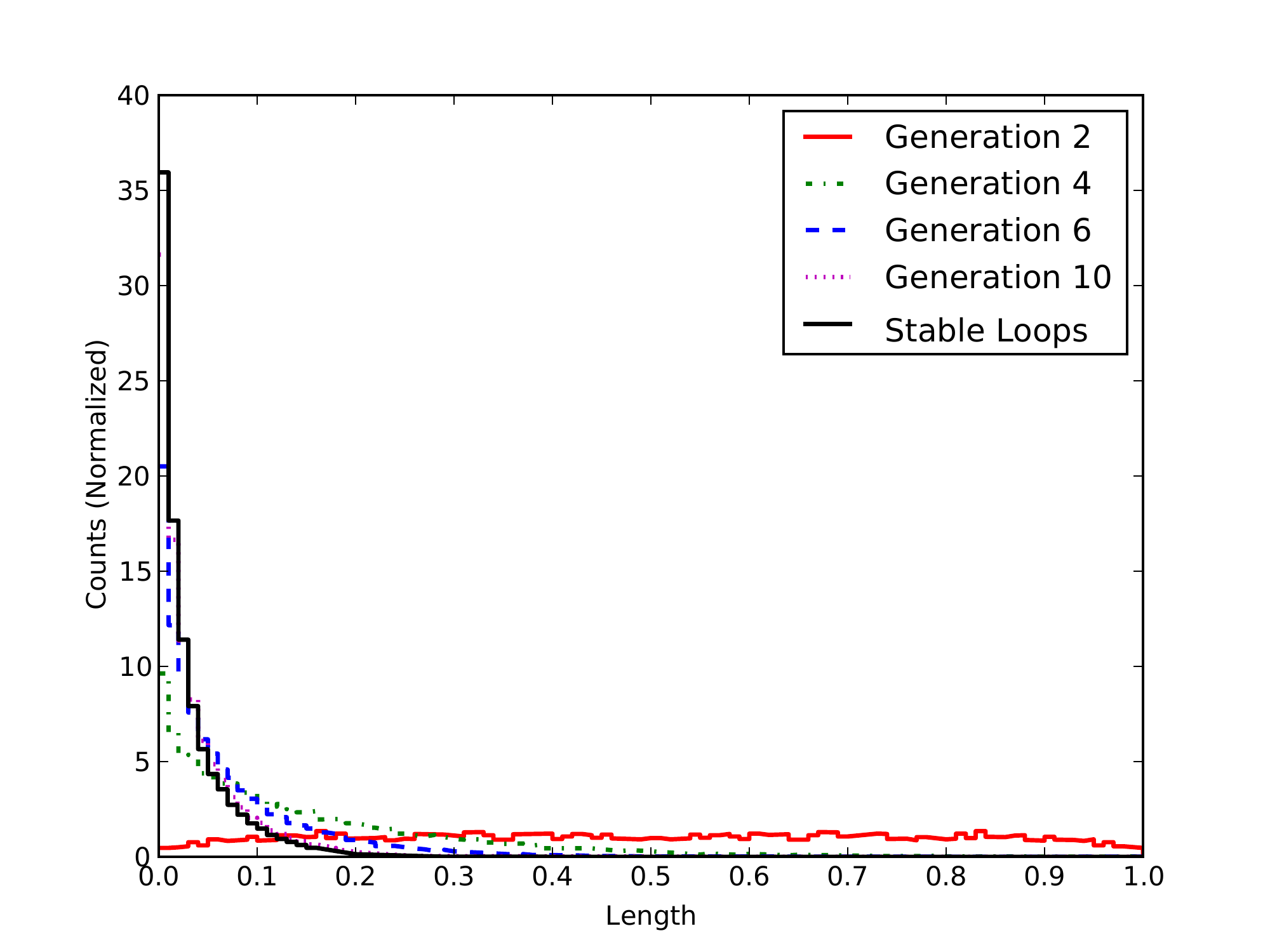}
  \caption{Length of loops versus generation. By generation 10 the
    distribution is nearly identical to the stable loop length
    distribution.  As fragmentation occurs the length of loops decreases.
    The distribution of the predominantly small length stable loops can be
    seen in Fig.~\ref{fig:stablelooplengths}.}
  \label{fig:looplengthplots}
\end{figure}

\begin{figure}
  \includegraphics[width=0.45\textwidth]{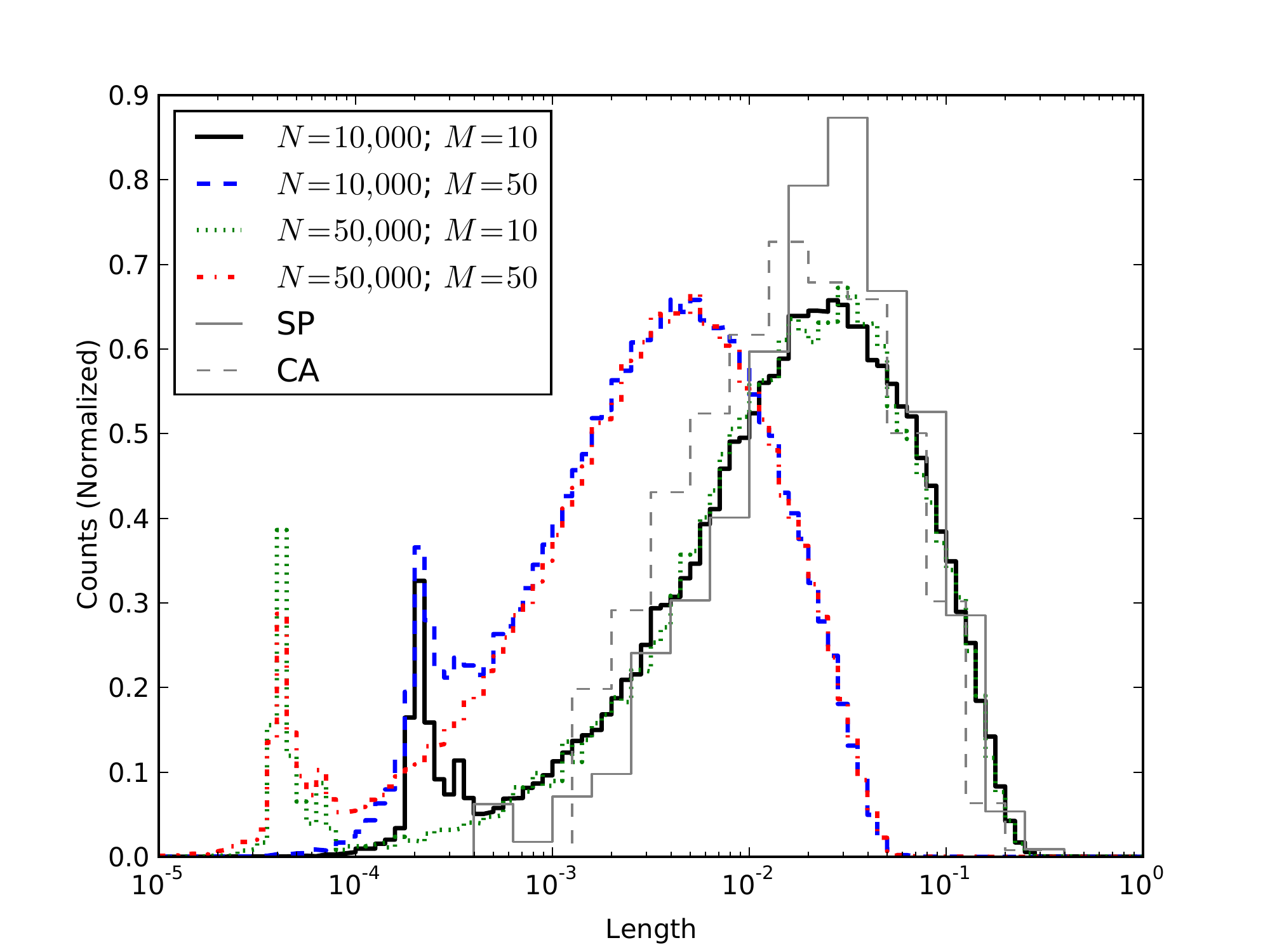}
  \caption{Length distribution of stable loops shown on a logarithmic scale
    to better see the small length behavior.  The resolution limit of the
    simulations is seen by the small peaks at $L\sim 2/N$. The length
    distribution is otherwise independent of resolution, $N$.  Comparison
    to the SP (solid, gray line) and CA (dashed, gray line) shows good
    general agreement.}
  \label{fig:stablelooplengths}
\end{figure}

Most of our data is collected for loops of 10,000 segments and 10 harmonic
modes; see Table~\ref{tab:loop-parameters}. In Fig.~\ref{fig:sampleloop} we
show a sample initial loop. We evolve
the loop as described in Sec.~\ref{sec:procedure} and keep track of the
fragmentations. We define loop ``generations'' in the following way. The
initial loop is the first generation.  When it splits into two loops, those
are in the second generation, and so on for successive generations. Once a
loop is in a stable configuration it is counted in each subsequent
generation.  For example, a stable loop that is generated in the third
generation will be counted in the third, fourth, \textit{etc.}\ generations.
At every generation, we can plot the length distribution of loops as shown
in Fig.~\ref{fig:looplengthplots}.  The plot shows that the second
generation has roughly a uniform distribution of lengths but that the
stable loop distribution is sharply peaked at small lengths.  The stable
loop length distributions are plotted in Fig.~\ref{fig:stablelooplengths}.
Though very small loops are formed through fragmentation the peaks in this
figure at $L\sim 2/N$ show the resolution limit of the simulations.  Clearly
the bulk of the stable loops are above our resolution limit.  For
comparison the SP and CA results are shown and are in good agreement with
our results. 

\begin{figure}
  \includegraphics[width=0.45\textwidth]{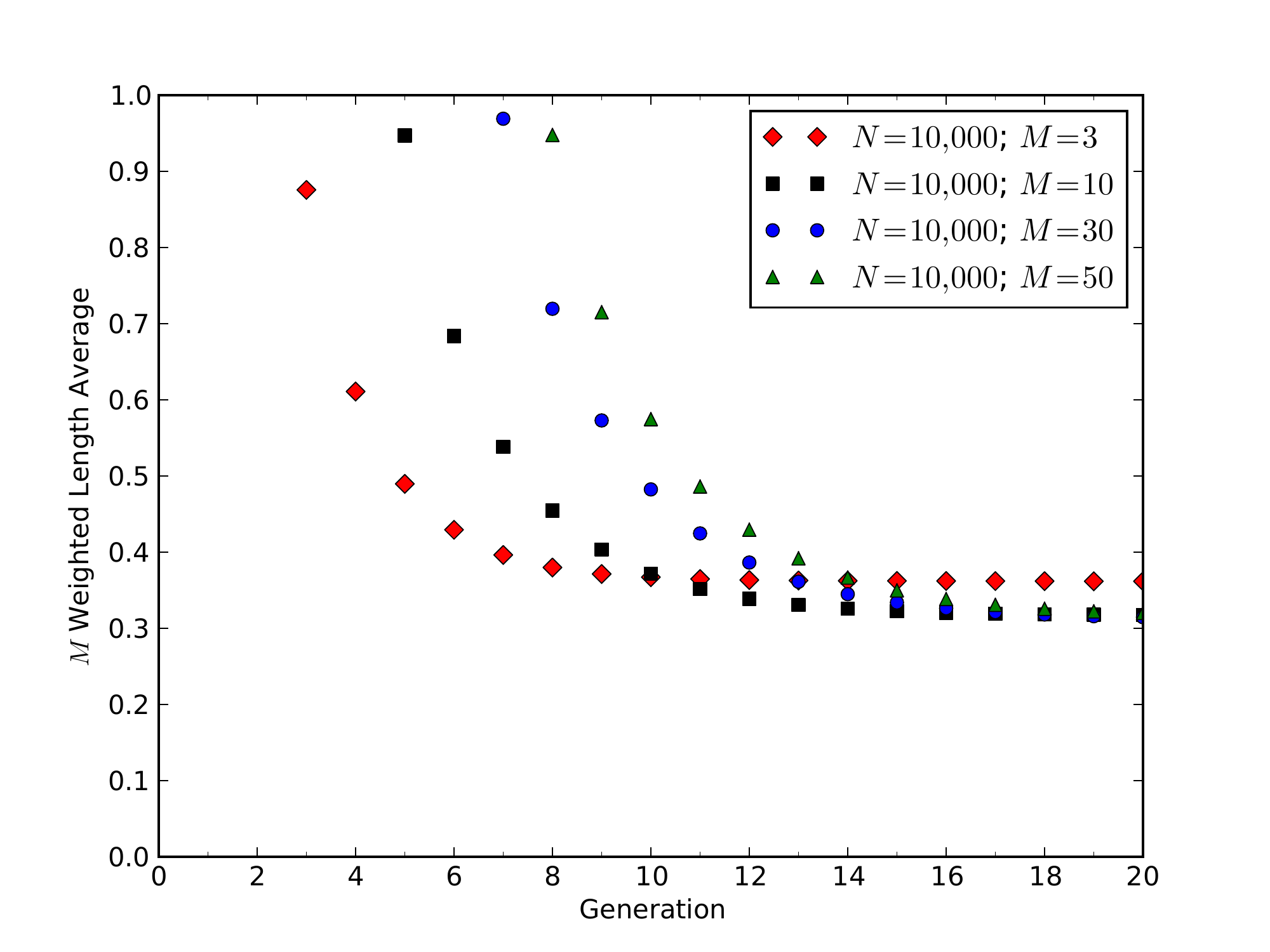}
  \caption{Average loop length multiplied by the number of harmonics in the
    parent loop versus generation. This weighted length asymptotes to
    $\sim0.33$ for all initial loops.}
  \label{fig:looplengthgenerationaverages}
\end{figure}

From the length distribution we calculate the average length per generation
for several values of $M$. In Fig.~\ref{fig:looplengthgenerationaverages} we
show the average length scaled by $M$ for the stable loops. All the plots
for $M=10, \ldots, 50$ asymptote to the same value, $0.33$, which clearly 
demonstrates that the average length of stable loops is proportional to 
$1/M$. The plot for $M=3$ asymptotes to $0.36$ and suggests that loops with
only very low harmonics could behave differently from loops with high
harmonics. However, even the $M=3$ value of $0.36$ is very close to
$0.33$ and we will disregard this difference in the discussion below.
Then the asymptote at $0.33 =1/3$ implies
that an initial loop with $M$ harmonics fragments into $3M$ stable
loops, as can also be read off from Table~\ref{tab:loop-parameters}.
Given that the initial loop is straight on the length scale $1/M$,
the stable loop is also composed of roughly straight segments of length
$\sim 1/3M$. This result solidifies and quantifies the finding in 
Ref.~\cite{Bennett:1987vf} that fragmentation does not continue indefinitely 
to smaller and smaller loops.

In terms of the ${\vec p}$ and ${\vec q}$ curves on
the KT sphere, since a loop with only the fundamental harmonic wraps
around a great circle once, we expect that stable loops will wrap 
the sphere only a third of the way. Also, the discussion below
shows that the ${\vec p}$ and ${\vec q}$ curves for the stable loops
occur in two disconnected segments on the KT sphere. Hence each segment
only covers a sixth of the KT sphere and can be thought of as an
arc on the KT sphere that extends $60^\circ$ on average. 

\begin{figure}
  \includegraphics[width=0.45\textwidth]{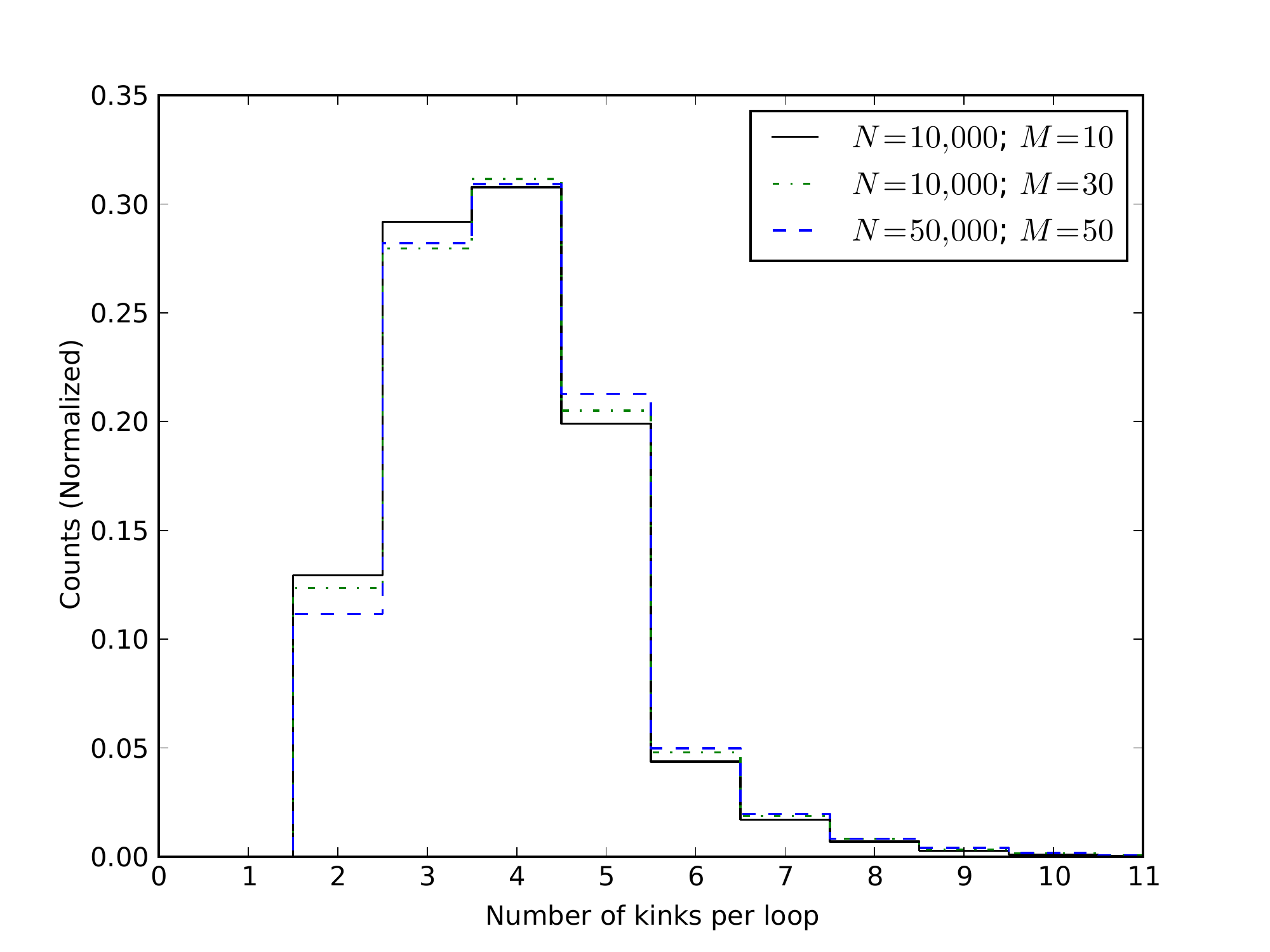}
  \caption{The number of kinks on stable loops. Each loop must have at
    least $2$ kinks. Most of the loops have between $2$ and $5$ kinks
    independent of resolution and the number of harmonic modes.}
  \label{fig:loopkinkcounts}
\end{figure}

Our result that the stable loops are made from roughly straight segments
can only work if the loops have kinks on them. It can be argued that the
average number of kinks should be four \cite{Casper1995}. Assume that there
is an initial loop with $N_k$ pre-existing kinks. (In our numerical work, the
initial loops were smooth and so $N_k=0$.) Every intercommutation event adds
four kinks and one extra loop to the system. Therefore after $n$
intercommutations there will be $4n+N_k$ kinks and $n+1$ loops. So the average
number of kinks per loop is $(4n+N_k)/(n+1)$ which goes to $4$ in the large
$n$ limit. The distribution of kinks, however, needs to be calculated
numerically and is shown in Fig.~\ref{fig:loopkinkcounts}. From this
argument and the numerical simulations we have the picture that stable
cosmic string loops have roughly straight sides and four corners (kinks) on
average.

\begin{figure}
  \includegraphics[width=0.45\textwidth]{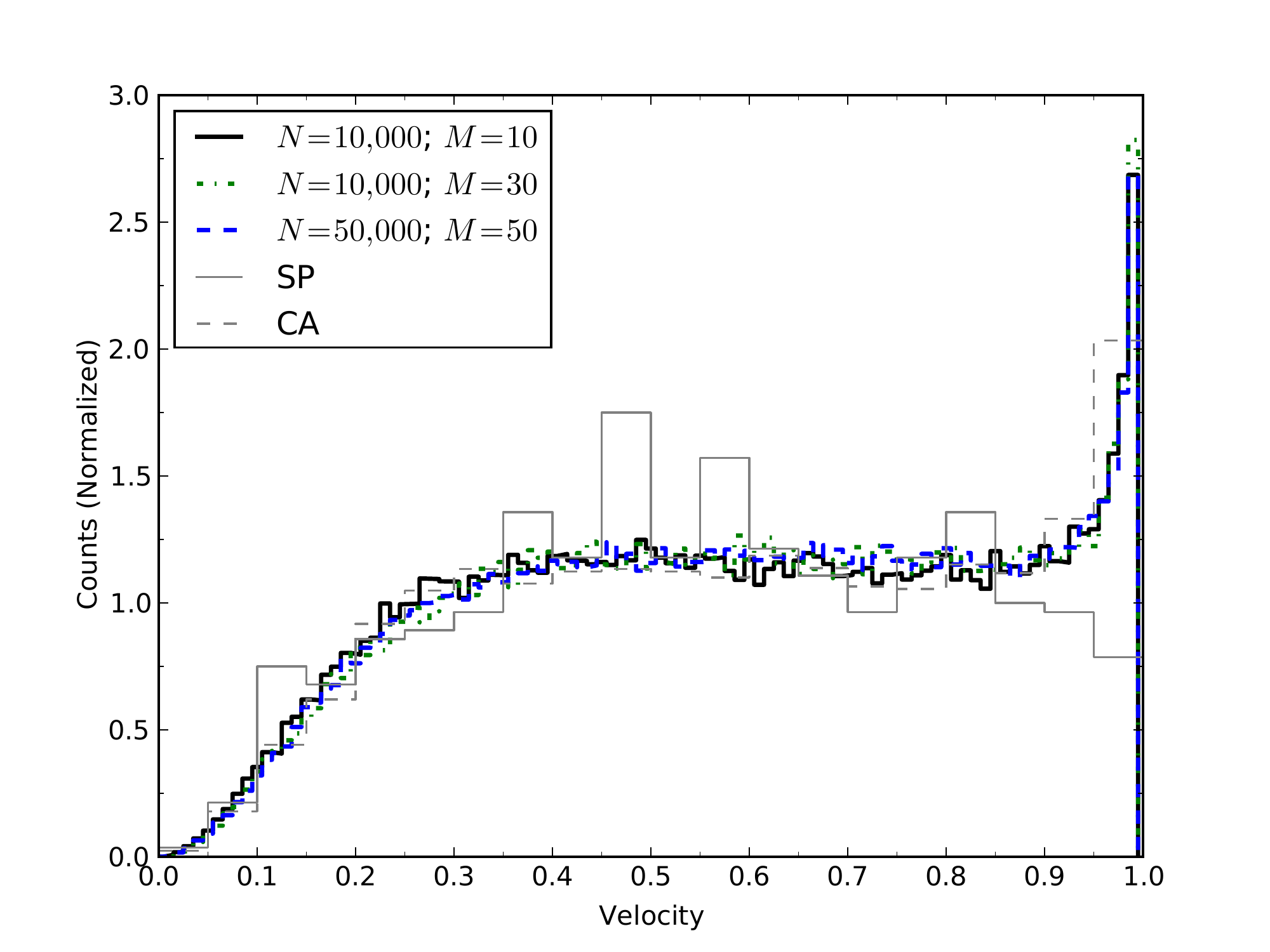}
  \caption{The velocity distribution of stable loops for various
    resolutions and numbers of modes.  The distribution is universal and
    sharply peaked at $v\sim 1$. Also shown are the SP (solid, gray line)
    and CA (dashed, gray line) for comparison.  The resolution of SP did
    not allow them to produce high velocity loops. Our results refine
    the peak at high velocity found by CA\@.}
  \label{fig:velocity-distribution}
\end{figure}

In general, stable loops have relativistic center of mass velocities as
shown in Fig.~\ref{fig:velocity-distribution}. To analyze the shape of a
stable loop, we must first transform to its center of mass frame. Care must
be taken to transform the coordinates so that the gauge conditions in
Eqs.~(\ref{eqn:constraints}) are satisfied as described in
Appendix~\ref{app:transformation}\@. Let us denote the resulting curves on
the KT-sphere in the rest frame by ${\vec P}$ and ${\vec Q}$.

\begin{figure}
  \includegraphics[width=0.45\textwidth]{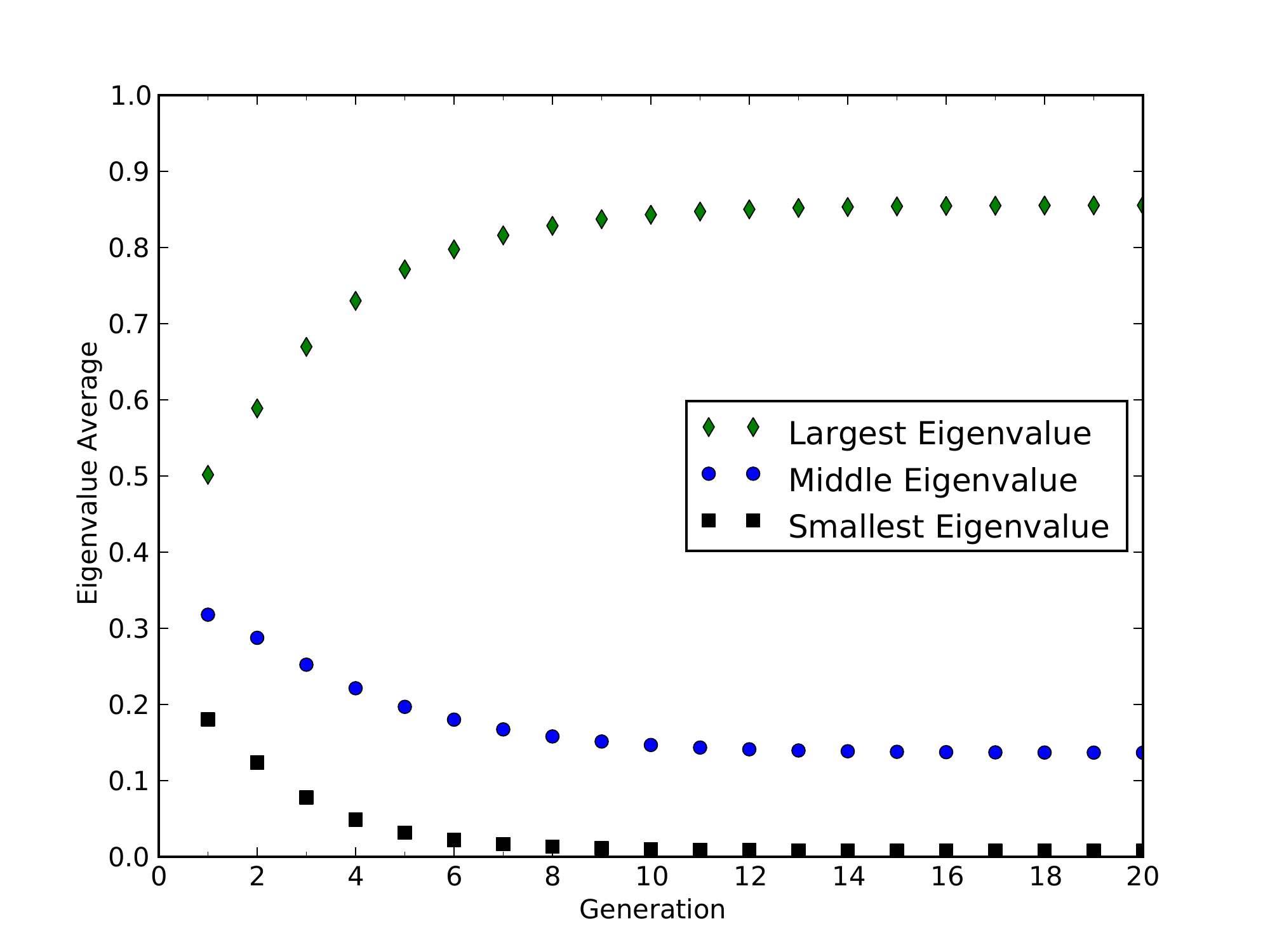}
  \caption{The average of the eigenvalues of the $\vec P$ and $\vec Q$
    ``moment of inertia'' tensors~(\ref{eqn:moment-of-inertia}) versus
    loop generation.  Since both $\vec P$ and $\vec Q$ have the same
    properties, both are included in the average to increase the
    statistics. We notice that the smaller eigenvalue approaches zero and
    the largest eigenvalue is much larger than the middle eigenvalue
    showing that the $\vec P$ and $\vec Q$ curves are localized on the
    KT-sphere. This plot is for $N=10,000$, $M=10$.  The asymptotic
    behavior is independent of resolution and the number of modes.}
    \label{fig:eigenvalue-average}
\end{figure}

To analyze the planarity of a stable loop, we calculate the ``moment of
inertia'' tensors for $\vec P$ and $\vec Q$ via
\begin{eqnarray}
  A_{ij} & = & \int P_i P_j\, d\sigma_- , \nonumber \\
  B_{ij} & = & \int Q_i Q_j\, d\sigma_+.
  \label{eqn:moment-of-inertia}
\end{eqnarray}
If the ${\vec P}$ (and similarly for the ${\vec Q}$) curve is uniformly
distributed on the KT-sphere, the three eigenvalues of the $A$ tensor would
be equal.  If one of the eigenvalues of $A$ vanishes, then ${\vec P}$ is
distributed in a plane; and if two eigenvalues vanish, the distribution is
lineal.  In Fig.~\ref{fig:eigenvalue-average} we show the average
eigenvalues as the loop continues to fragment. It is clear that one of the
eigenvalues vanishes, and the largest eigenvalue is much larger than the
middle eigenvalue. So the ${\vec P}$ curve is mostly in one direction on
the KT-sphere (say around the $z$-axis), with a little spread in some other
direction. Since the centroid of the ${\vec P}$ curve has to vanish in the
center of mass frame (see the discussion below Eq.~(\ref{eqn:closure})),
the ${\vec P}$ curve must contain two short segments, say one near the
north pole and the other near the south pole. In an idealized
(``degenerate'') case, the ${\vec P}$ curve would consist of just 2 points
on the KT-sphere. Similarly the ${\vec Q}$ curve would consist of just 2
other points. Since the loop is a sum of the left- and right-movers, this
implies that the loop is planar and defined by the plane of the ${\vec P}$
and ${\vec Q}$ curves.

\begin{figure}
  \includegraphics[width=0.45\textwidth]{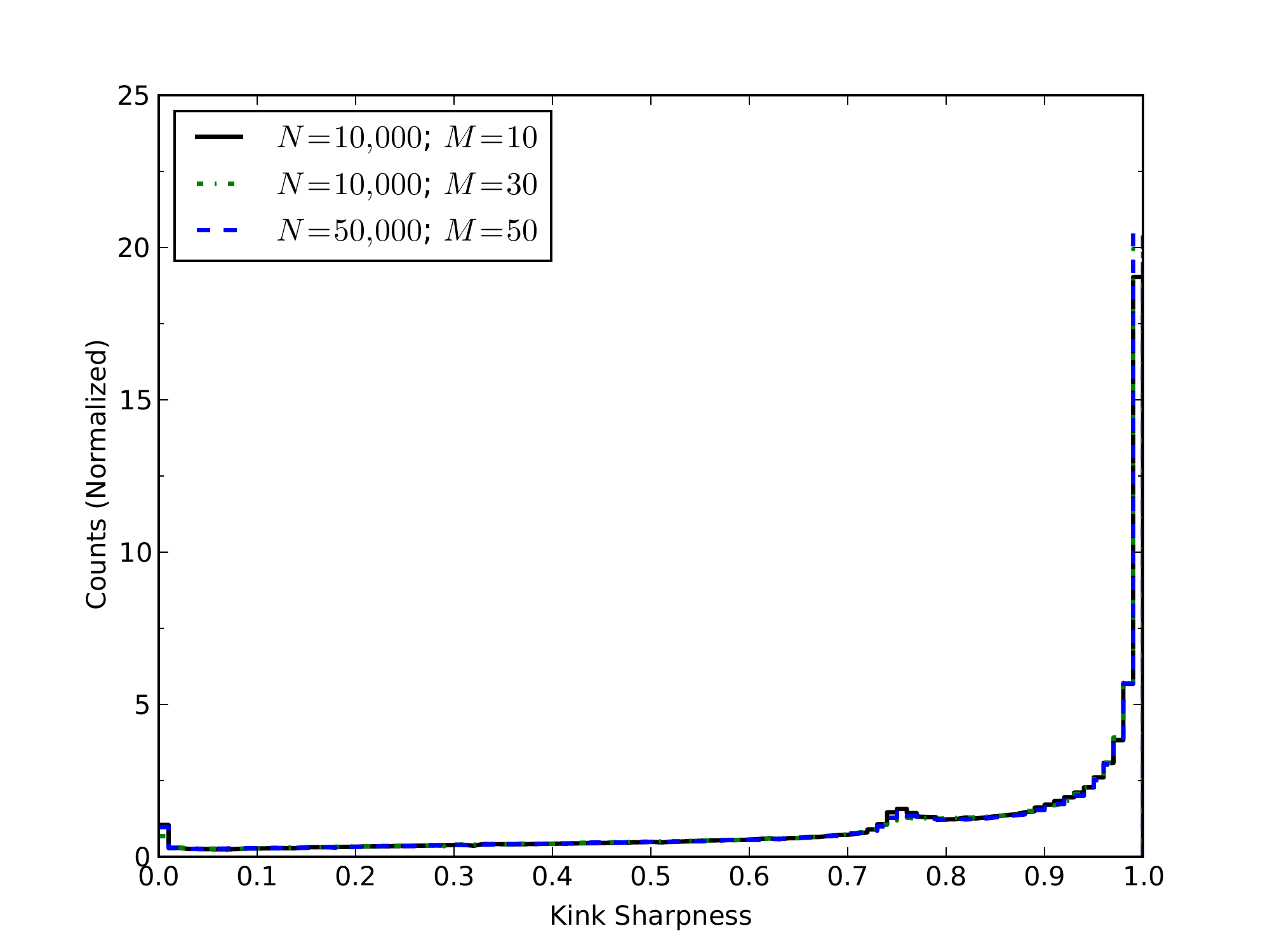}
  \caption{The kink sharpness distribution (see
    Eq.~\ref{eq:kinksharpness}).  The distribution is independent of
    resolution and number of modes on the initial string.  We see that the
    distribution is sharply peaked at $1$, that is where the $\vec P$
    before and after the kink are anti-parallel.}
  \label{fig:kinksharpnesshistograms}
\end{figure}

As another check of this picture, we have plotted the kink sharpness for
the ${\vec p}$ curves, defined as \cite{Copeland2009}
\begin{equation}
  \psi_a = \frac{1}{2} (1- {\vec P}_- \cdot {\vec P}_+ ),
  \label{eq:kinksharpness}
\end{equation}
where ${\vec P}_\pm$ are the values of ${\vec P}$ on either side of the
kink. The plot in Fig.~\ref{fig:kinksharpnesshistograms} demonstrates a
peak near sharpness of 1, which implies ${\vec P}_- \approx - {\vec
  P}_+$. This shows that there are 180 degree jumps on the KT-sphere, in
accordance with our picture that the ${\vec P}$ curve corresponds to two
small anti-podal regions on the KT-sphere.

\begin{figure}
\includegraphics[width=0.45\textwidth]{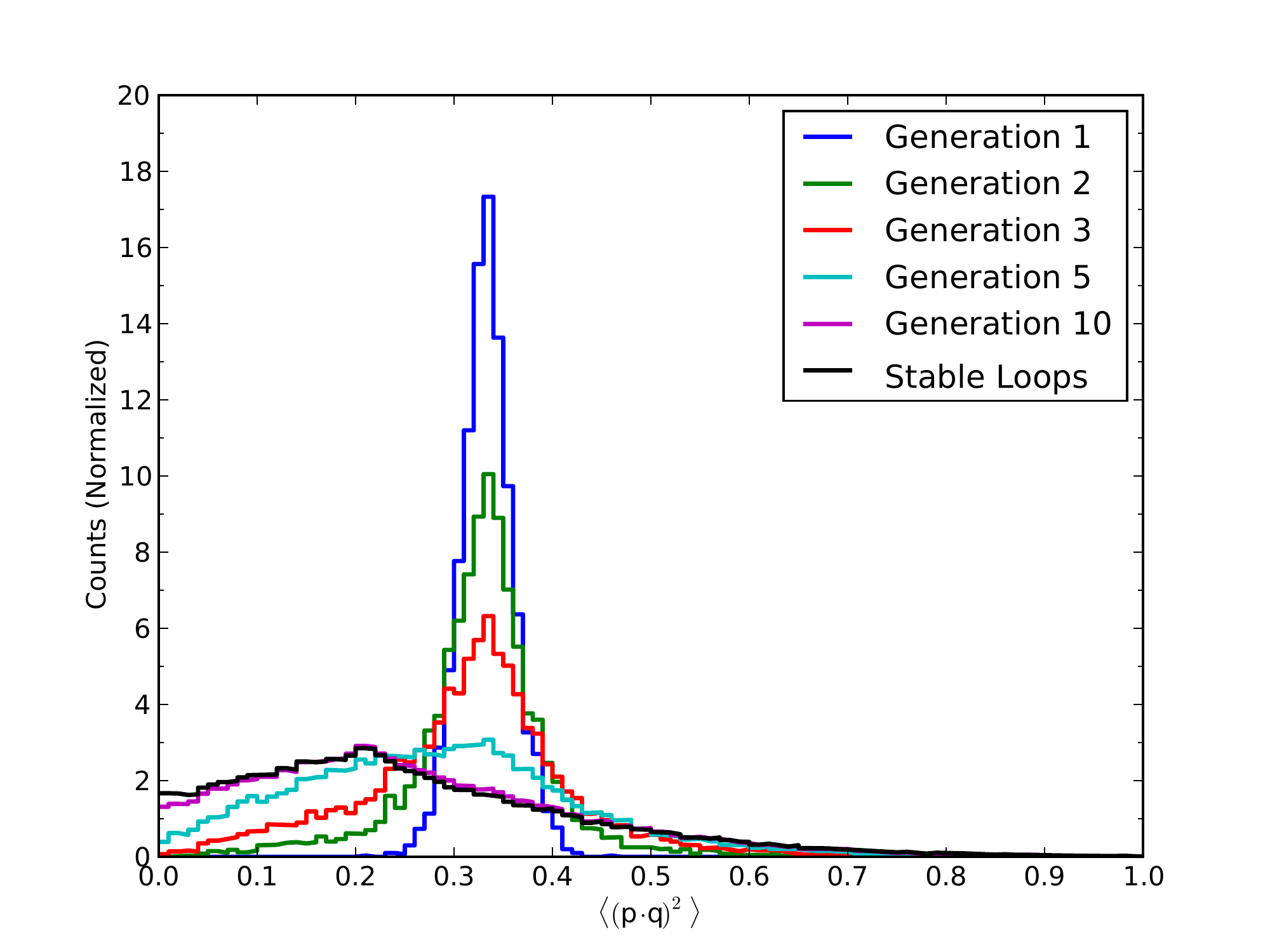}
  \caption{The distribution of $\langle({\vec P}\cdot {\vec Q})^2\rangle$
    showing that the left- and right- movers become more orthogonal to each
    other with more fragmentation.}
  \label{fig:pdotqdistn}
\end{figure}

We can go a bit further and ask for correlations between the left- and
right-movers. For this we plot the distribution of 
$\langle({\vec P}\cdot {\vec Q})^2\rangle$ in Fig.~\ref{fig:pdotqdistn}. 
(Angular brackets denote average over a loop.)
This plot shows that ${\vec P}$ and ${\vec Q}$ tend to get more
orthogonal with fragmentation.

This picture of stable loops implies that the ${\vec P}$ and $-{\vec Q}$
curves rarely cross each other and hence that cusps should be suppressed.
For an initial loop with $M$ modes we expect it to contain approximately
$M^2$ cusps. (Each mode roughly corresponds to a great circle on the
KT-sphere so the number of intersections is proportional to $M^2$.)

\begin{figure}
  \includegraphics[width=0.45\textwidth]{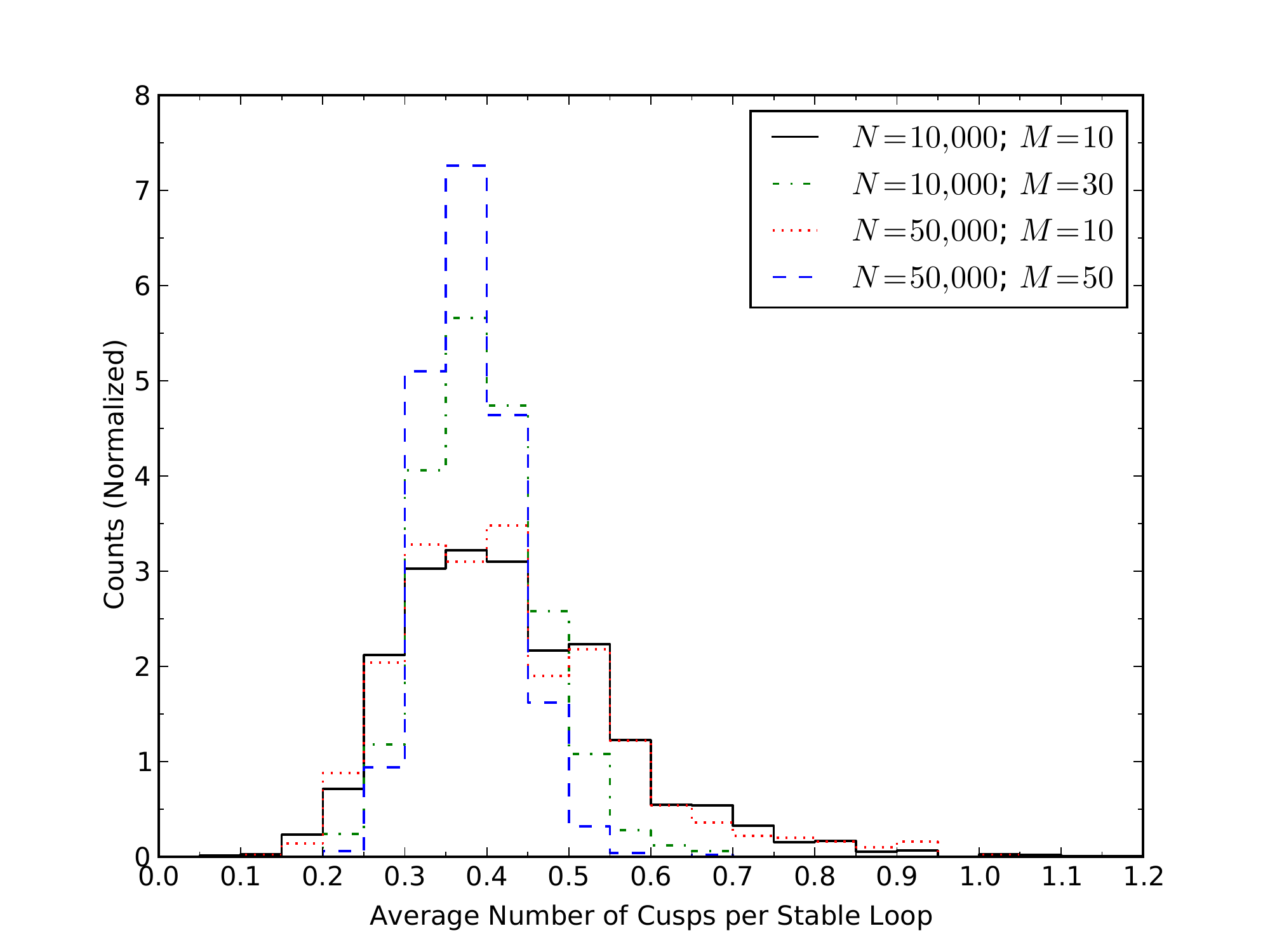}
  \caption{The average number of cusps on each stable loop.  On average we
    see that there is approximately a $40\%$ chance of a stable loop to
    contain a cusp.  This result is independent of resolution and the
    number of modes on the initial loop.}
  \label{fig:average-stable-cusps}
\end{figure}

We have seen that on average $3M$ stable loops are produced. If $f$ is the
``cusp survival fraction'' --- the fraction of cusps on the initial loop
that survive on the stable loops --- then on average we expect $M^2 f /3M$
cusps on each loop.  Fig.~\ref{fig:average-stable-cusps} shows that, on
average, each stable loop has a $40\%$ of containing a cusp independent of
resolution and the number of modes on the initial loop so the cusp survival
fraction is found to be $f\approx 6/5M$. (We have verified this is true in
our simulations but do not show the results here.)

\section{Length dependent results}
\label{sec:lwa}

The focus of this work is on the average properties of all stable loops.
The statistics we have presented give equal weights to all loops. For some
observational signatures, however, long loops may be more relevant than
small loops. Hence, in this section, we provide some statistics as a
function of loop length.  The results shown here are for the $N=10,000$ and
$M=10$ runs.  Logarithmic length bins are chosen based on the distribution
of loop lengths (see Fig.~\ref{fig:stablelooplengths}).  Ten bins are
chosen between lengths of $10^{-3}$ and $0.1$.  The initial bin begins at a
length of $5\times10^{-4}$ to ensure that all loops are above our
resolution limit; approximately $94\%$ of the loops are included.  The
final bin includes all loops with lengths longer than $0.1$.

\begin{figure}
  \includegraphics[width=0.45\textwidth]{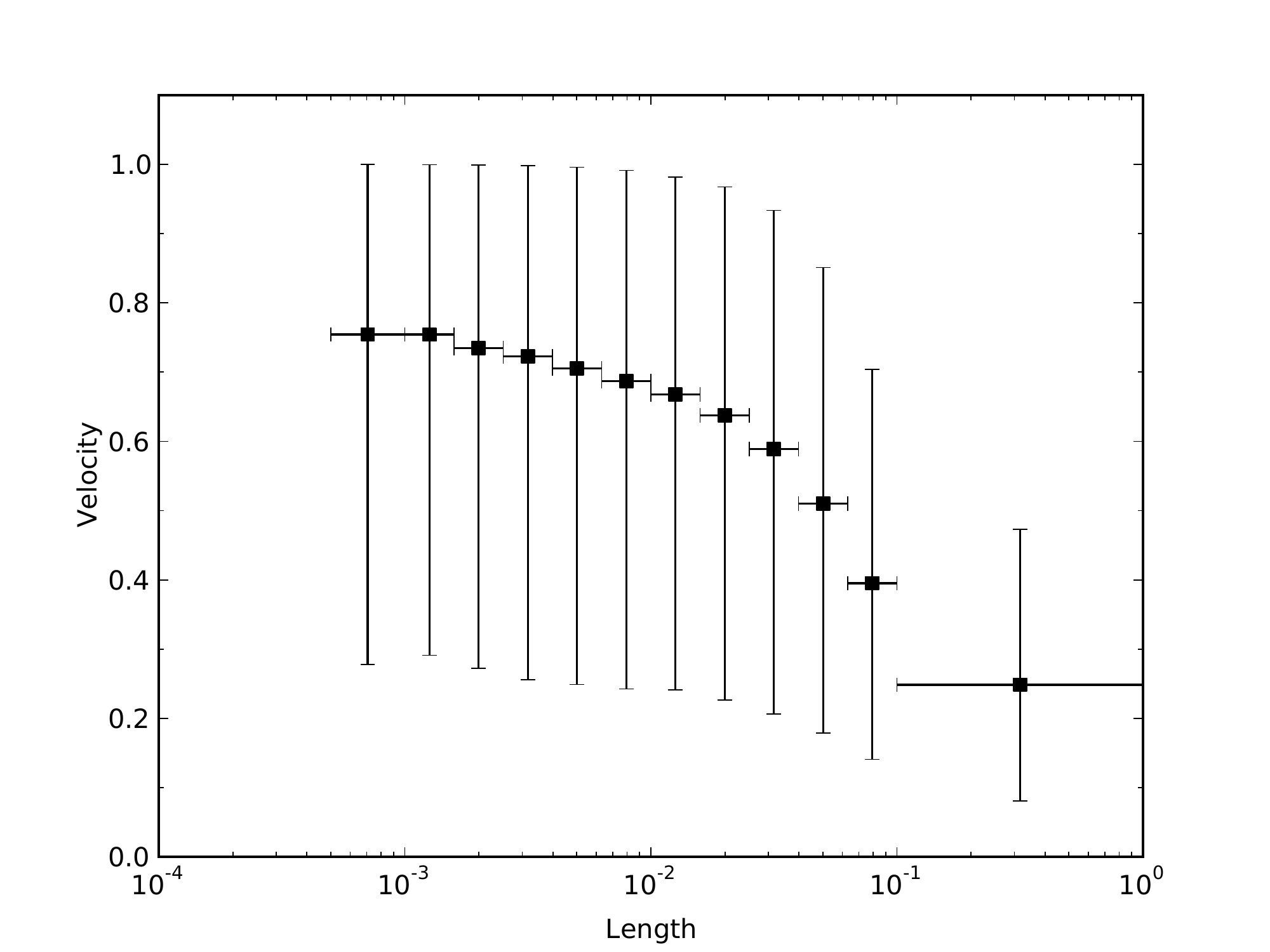}
  \caption{The velocity of stable loops versus the length of the loop. The
    $x$ error bars show the width of the length bins employed and the $y$
    error bars show the $95$ percentile ranges in each length bin. The
    square symbol shows the average in the bin.  Short loops have center of
    mass velocity $\sim 1/\sqrt{2}$ which is the root-mean-squared velocity
    of the string in the initial loop. Longer loops tend to have a lower
    velocity consistent with the fact that the initial loop is at rest.}
  \label{fig:velocityvslength}
\end{figure}

\begin{figure}
  \includegraphics[width=0.45\textwidth]{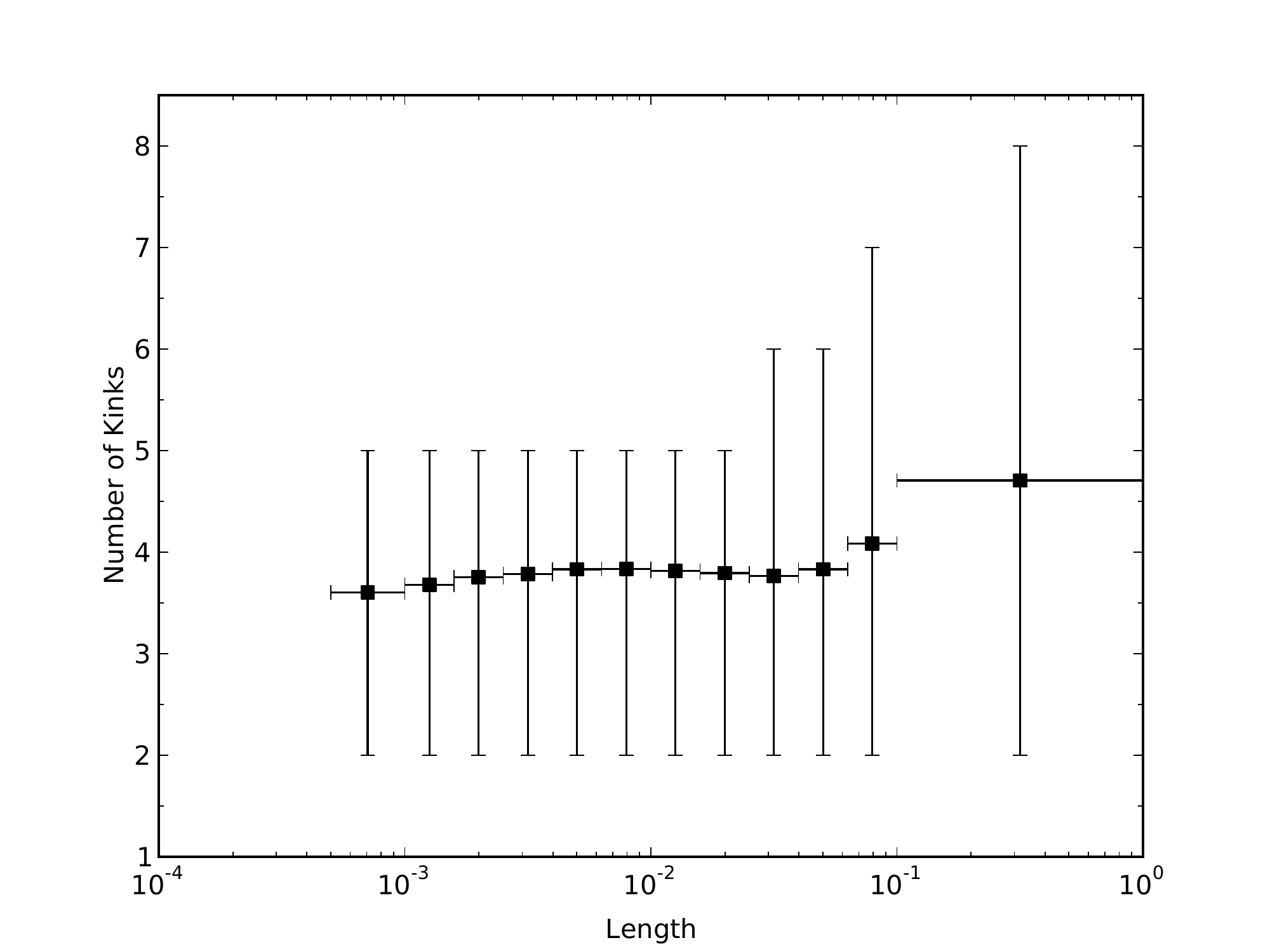}
  \caption{Similar to Fig.~\ref{fig:velocityvslength} now for the number of
    kinks on stable loops versus the length of the loop. Longer loops tend
    to have a larger number of kinks though the variation is slight. The
    loop with the maximum number of kinks has 8 kinks instead of the
    canonical 4.}
  \label{fig:kinksvslength}
\end{figure}

\begin{figure}
  \includegraphics[width=0.45\textwidth]{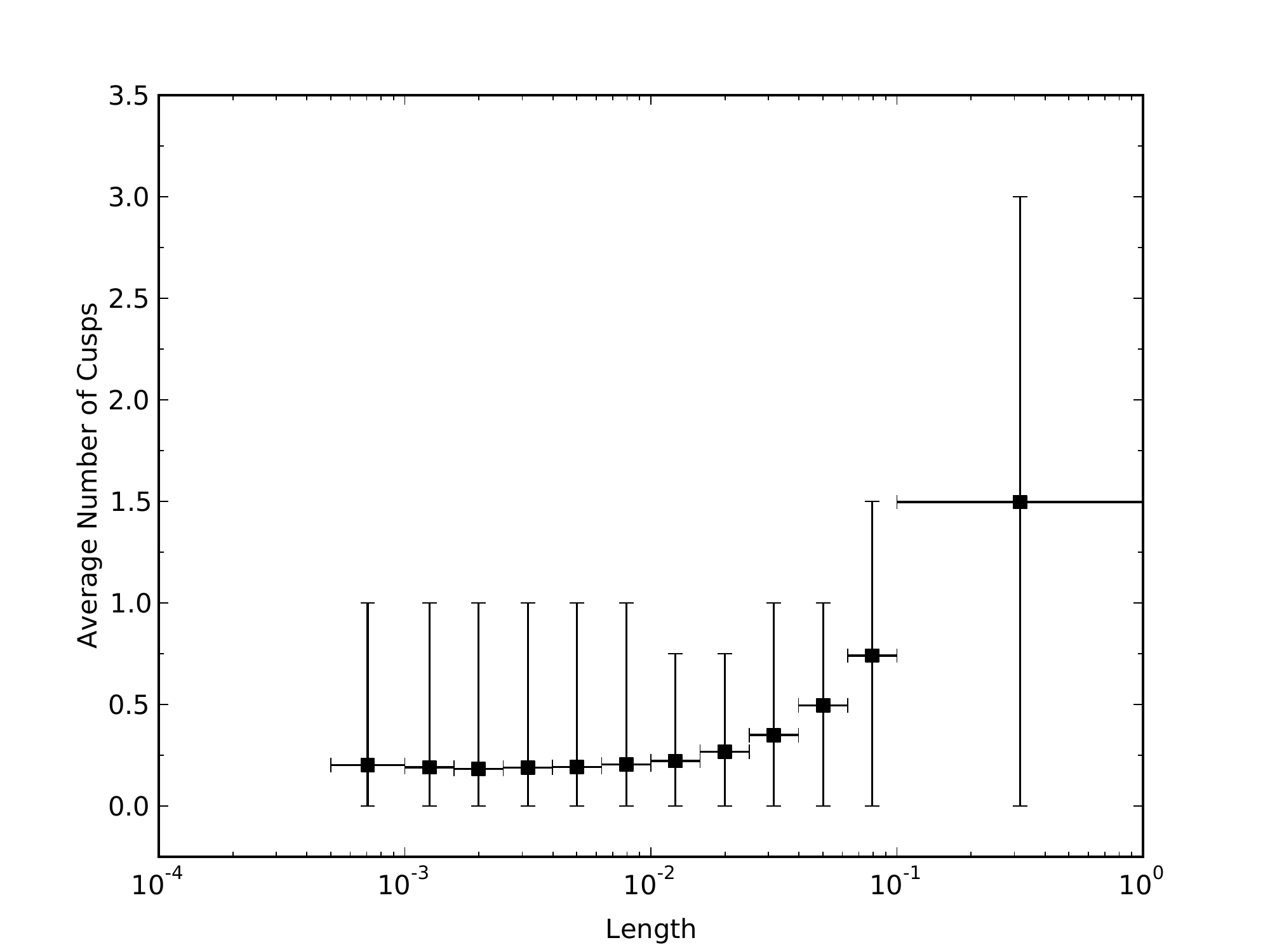}
  \caption{Similar to Fig.~\ref{fig:velocityvslength} now for the average
    number of cusps on stable loops versus length of the loop. Longer loops
    typically contain more cusps on average. The loop with the maximum
    number of cusps has only 3 cusps.}
  \label{fig:cuspsvslength}
\end{figure}

\begin{figure}
  \includegraphics[width=0.45\textwidth]{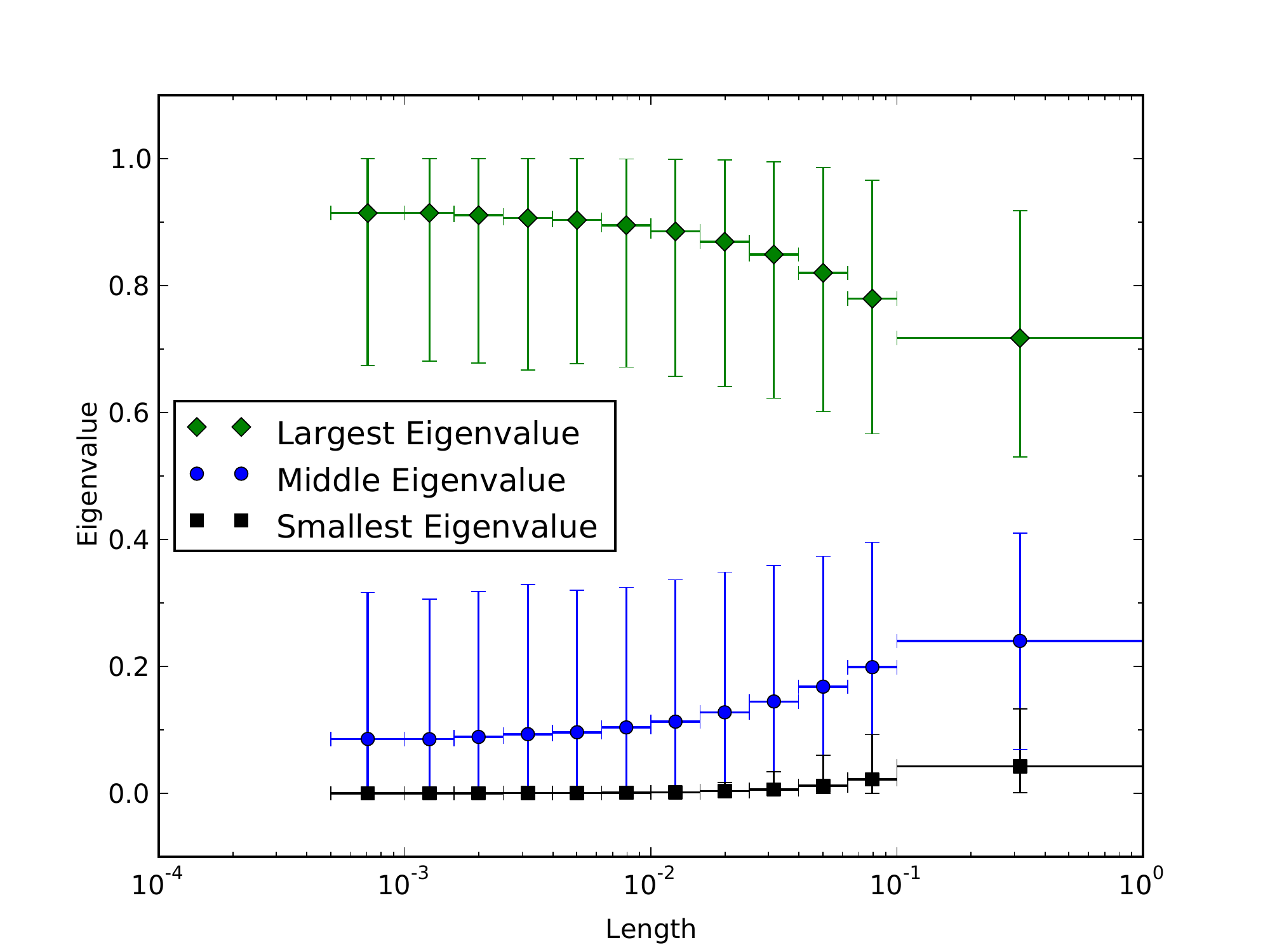}
  \caption{Similar to Fig.~\ref{fig:velocityvslength} now for the
    eigenvalues of the moment of inertia tensors of the stable loops versus
    length. Longer loops are also planar and only slightly less so than
    smaller loops.}
  \label{fig:eigenvaluesvslength}
\end{figure}

The results are shown in
Figs.~\ref{fig:velocityvslength}--\ref{fig:eigenvaluesvslength}.  These
figures show the velocity, number of kinks, number of cusps, and
eigenvalues, respectively, for each length bin.  In the plots the $x$ error
bar shows the width of the bin and the $y$ error bar provides the $95$
percentile range for the values.  The results are as expected.  The
smallest loops are predominantly high velocity, contain approximately four kinks
per loop, have a small chance of containing a cusp, and are planar.  The
largest loops are predominantly lower velocity, contain more than four
kinks, have a higher chance of containing a cusp, and are less planar.

\section{Discussion}
\label{sec:discussion}

Our numerical results point to the picture that stable loops are
deformations of degenerate kinky loops. In this section we provide further
evidence for this picture by demonstrating that there exists a class of
perturbed degenerate kinky loops that are stable.

\begin{figure}
  \scalebox{0.6}{\includegraphics[angle=-90]{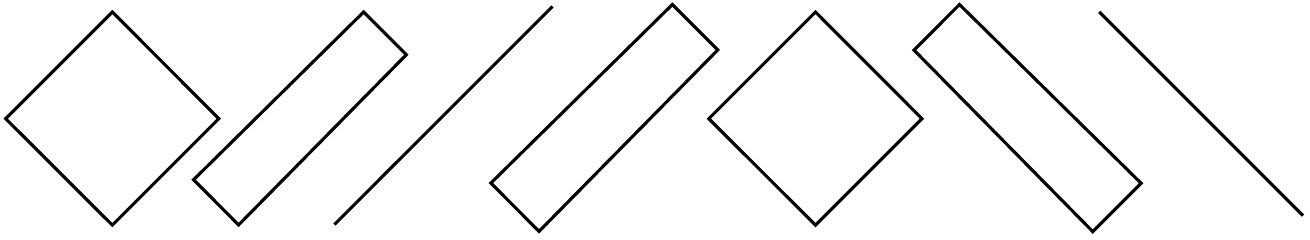}}
  \caption{Snapshots during the evolution of the degenerate kinky loop. Note
    that the loop periodically collapses to a double line.}
  \label{fig:dkloop}
\end{figure}

We begin by describing the degenerate kinky loop
\cite{Garfinkle1987},
\begin{eqnarray}
  {\vec p}_{\mathrm{dk}} &=& \cos (\alpha_0) {\hat z}, \nonumber \\
  {\vec q}_{\mathrm{dk}} &=& \cos (\beta_0) {\hat x},
  \label{eq:degenerate-kink}
\end{eqnarray}
where
\begin{equation}
  \alpha_0 = \pi \lfloor 2\sigma_-\rfloor , \quad
  \beta_0 = \pi  \lfloor 2\sigma_+ \rfloor ,
\end{equation}
$\lfloor x\rfloor$ denotes the greatest integer less than or equal to $x$,
and $\sigma_\pm\in[0,1]$.  In Fig.~\ref{fig:dkloop} we show some snapshots
of a degenerate kinky loop.

\begin{figure}
  \includegraphics[height=1.25in]{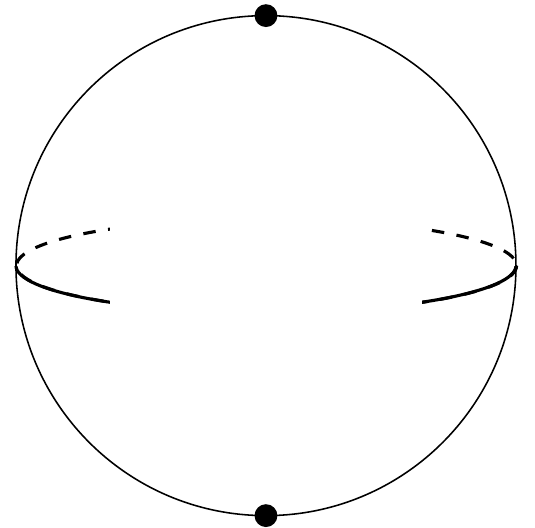}
  \caption{The $\vec p$ and $\vec q$ curves for a perturbed degenerate
    kinky loop~(\ref{eq:perturbed-dk}) plotted on the KT-sphere. The $\vec p$
    curve is represented by the two points at the poles.  The $\vec q$
    curve is represented by the arcs along the equator.}
  \label{fig:pertdkloop}
\end{figure}

Note that the degenerate kinky loop collapses to a double line twice during
an oscillation period, yet the stable loops do not self-intersect. This
must be due to the fact that stable loops are not exactly degenerate. To
show that this is the case, consider
\begin{eqnarray}
  {\vec p} &=& \cos (\alpha) {\hat z},  \nonumber \\
  {\vec q} &=& \cos (2\pi\epsilon \sigma_+ + \beta) {\hat x}
  + \sin (2\pi\epsilon \sigma_+ + \beta) {\hat y}
  \label{eq:perturbed-dk}
\end{eqnarray}
where $0 < \epsilon < 1$ and now
\begin{eqnarray}
  \alpha &=& \pi \lfloor 2\sigma_- \rfloor, \nonumber \\ 
  \beta  &=& (1-\epsilon )\pi \lfloor 2\sigma_+ \rfloor .
\end{eqnarray}
As illustrated in Fig.~\ref{fig:pertdkloop},
these perturbations stretch out the point corresponding to ${\vec q}$
on the KT-sphere for the degenerate kinky loop to an arc of length
$\epsilon \pi$ \cite{Garfinkle1987}.

We can check explicitly that the self-intersection condition
\begin{equation}
  {\vec x}(\sigma,t) = {\vec x}(\sigma',t)
\end{equation}
is satisfied for such a loop if and only if $\sigma = \sigma'$ (mod
$1$), for any $t$, that is, the loop does not intersect at any time. We
first write the intersection condition in terms of ${\vec a}$ and ${\vec b}$,
\begin{equation}
  {\vec a}(\sigma-t) + {\vec b}(\sigma+t) =
  {\vec a}(\sigma'-t) + {\vec b}(\sigma'+t).
\end{equation}
Since for our loop~(\ref{eq:perturbed-dk}) we have ${\vec p}\cdot {\vec q}
=0$ for all $\sigma$ and $t$, we also have ${\vec a}\cdot {\vec b} =0$, and
self-intersection requires
\begin{eqnarray}
  {\vec a}(\sigma-t) & = & {\vec a}(\sigma'-t), \nonumber \\
  {\vec b}(\sigma+t) & = & {\vec b}(\sigma'+t).
\end{eqnarray}
The first condition is satisfied for $\sigma_-=\sigma_-'$ or for
$\sigma_-' = 1 - \sigma_-$. Only the latter possibility is of interest 
since we are looking for an intersection with $\sigma \ne \sigma'$. 
The second condition for the $x$ and $y$ components leads to
\begin{eqnarray}
  \sin (2\epsilon\pi \sigma_+)&=& \sin (\epsilon \pi) -
  \sin (2\epsilon\pi \sigma_+' - \epsilon \pi )  \nonumber \\
  \cos (2\epsilon\pi \sigma_+ )&=&
  1+\cos(\epsilon\pi) - \cos (2\epsilon\pi \sigma_+' - \epsilon \pi ) 
  \label{eq:dk-intersection}
\end{eqnarray}
Note that \textit{both} equations must be satisfied for a self-intersection
to occur.

With the help of some trigonometric identities, we can rewrite
(\ref{eq:dk-intersection}) as
\begin{eqnarray}
  \sin\left[\epsilon\pi \left(\sigma_+ +\sigma_+'- \frac12 \right) \right]
  \cos\left[\epsilon\pi \left(\sigma_+ -\sigma_+'+\frac12 \right) \right]
  && \nonumber \\
  && \hskip -1.5 in = \sin\left (\frac{\epsilon\pi}{2}\right )
  \cos\left (\frac{\epsilon\pi}{2}\right ) ,\nonumber \\
  \cos\left [\epsilon\pi \left(\sigma_+ +\sigma_+'- \frac12 \right) \right ]
  \cos\left [\epsilon\pi \left(\sigma_+ -\sigma_+'+\frac12 \right) \right ]
  && \nonumber \\
  && \hskip -1.5 in 
  = \cos^2\left (\frac{\epsilon\pi}{2}\right ) .
  \label{eq:4cosines}
\end{eqnarray}
Taking the ratio of these equations leads to
\begin{equation}
  \tan\left[\epsilon\pi \left(\sigma_+ +\sigma_+'-\frac12 \right) \right]
  = \tan\left( \frac{\epsilon\pi}{2} \right),
\end{equation}
so that $\sigma_+' +\sigma_+ =1$.
By squaring and adding the equations, we obtain
\begin{equation}
  \cos\left [\epsilon\pi \left(\sigma_+ -\sigma_+'+\frac12\right) \right ]
  = \pm \cos\left (\frac{\epsilon\pi}{2}\right ) .
\end{equation}
so that the second relation in Eq.~(\ref{eq:4cosines}) leads to
\begin{equation}
  \cos\left [\epsilon\pi \left(\sigma_+ -\sigma_+'+\frac12\right) \right ]
  = 
  \cos\left [\epsilon\pi \left(\sigma_+ +\sigma_+'-\frac12\right) \right ]
\end{equation}
or
\begin{equation}
\sin\left(\epsilon\pi \sigma_+ \right)
\sin\left[\epsilon\pi \left ( \sigma_+' -\frac12 \right ) \right ] =0
\end{equation}
which implies $\sigma_+ = 0$ or else $\sigma_+' =1/2$. 
(Solutions such as $\sigma_+ = 1 /\epsilon$ can be ignored because
we restrict $\sigma_+$ to lie in the interval $[0,2\pi]$ and
$|\epsilon|<1$.)  Therefore the only solution is of the type $\sigma_+ =0$,
$\sigma_+'=1$ or else $\sigma_+'=1/2$, $\sigma_+=1/2$, which are both
trivial.  This shows that there are no self-intersections. The conclusion
is intuitive since ${\vec b}$ is obtained by integrating ${\vec q}'$ and
hence is a vector that lies in the $xy-$plane (see
Fig.~\ref{fig:pertdkloop}) and rotates in this plane. Since $0< \epsilon <
1$, ${\vec b}$ never attains the same vector value twice for $\sigma_+ \in
[0,1]$.  

\begin{figure}
  \includegraphics[height=2.5in]{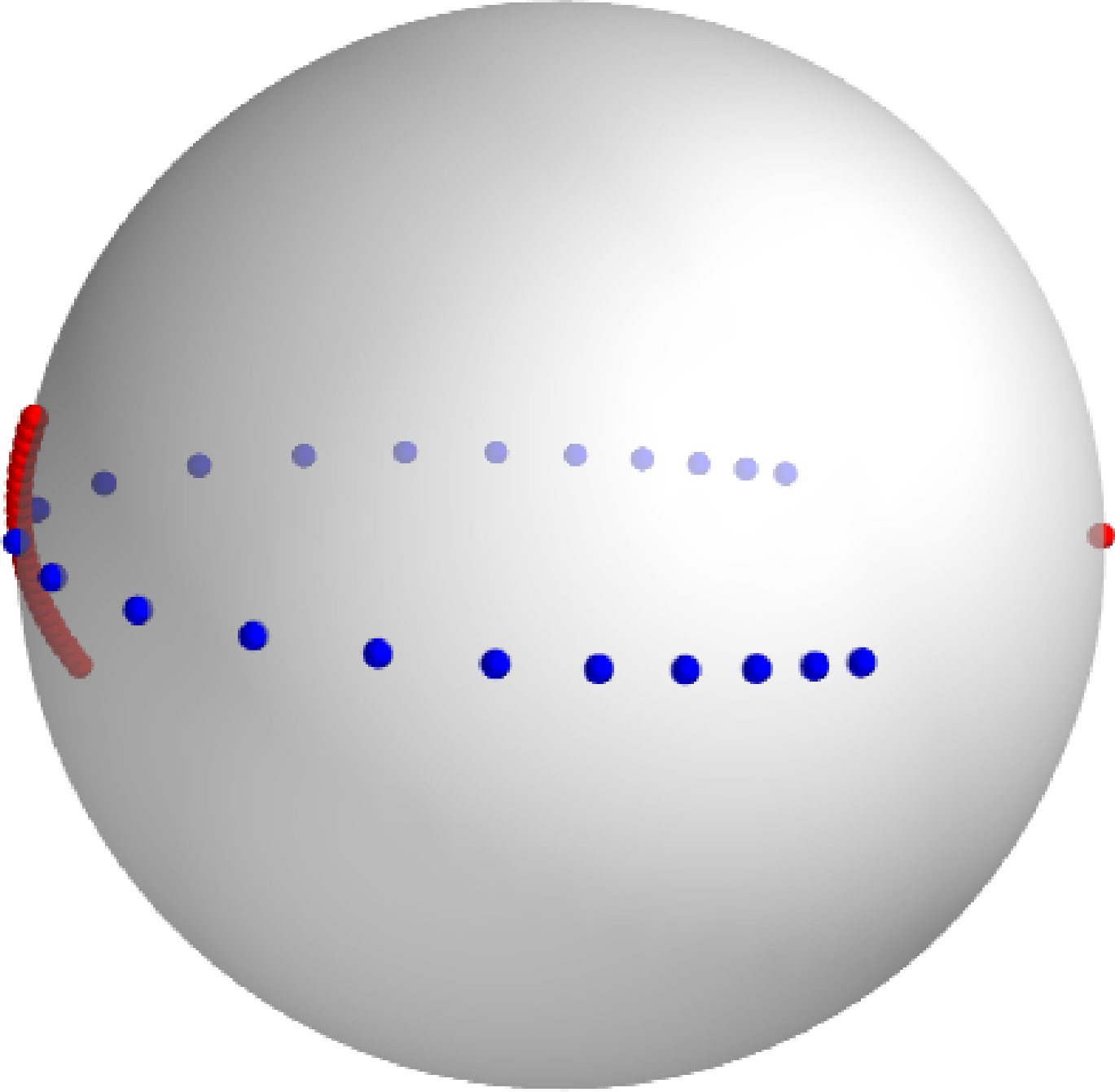}
  \caption{The ${\vec P}$ and $-{\vec Q}$ curves plotted on the KT-sphere
    for a stable loop from our simulations.  This loop contains three kinks
    and one cusp.  The two curves are represented by the dots.  The $\vec
    Q$ curve is the dots along the equator.  The $\vec P$ curve is the dots
    perpendicular to the equator (running vertically on the left of the
    sphere) and includes the lone anti-podal point (on the right of the
    sphere). Note that the points on the sphere are not weighted
    equally and the centroids of both $\vec P$ and $\vec Q$ curves
    are at the center of the sphere.}
  \label{fig:stablecuspyloop}
\end{figure}

The perturbed degenerate kinky loop discussed above does not contain any
cusps while our numerical results show that roughly half of the stable
loops have a cusp. To understand the ${\vec P}$ and ${\vec Q}$ curves of a
stable loop with cusps we show an example of these curves for a sample loop
from our numerical runs in Fig.~\ref{fig:stablecuspyloop} which contains
three kinks and one cusp.  In this example the $\vec P$ curve is like our
perturbed degenerate kinky loop but the $\vec Q$ curve is much more elongated.
The two curves are in planes nearly perpendicular to each other.

To construct a stable loop with cusps consider
\begin{eqnarray}
  \label{eq:pqcuspy}
  {\vec p} &=& \sin(\beta_-)\hat x +\cos(\beta_-)\hat z, \\
  {\vec q} &=& \sin(\beta_+)\left[\sin(\phi_+)\hat x +
    \cos(\phi_+)\hat y\right] - \cos(\beta_+)\hat z,
  \nonumber
\end{eqnarray}
where
\begin{equation}
  \beta_\pm = 2\pi\epsilon_\pm\sigma_\pm + (1-\epsilon_\pm)\pi\lfloor
	  2\sigma_\pm\rfloor,
\end{equation}
and $\phi_+$ is an arbitrary phase.  There are two cusps due to the
intersections at $\pm {\hat z}$.  These occur at $\sigma=t=0$ and
$\sigma=t=1/2$.  We have numerically confirmed that this loop is stable for
the parameter choice
\begin{equation}
  \epsilon_- = \frac{1}{\sqrt{58}}, \quad \epsilon_+=\frac{1}{\sqrt{95}},
  \quad \phi_+=\frac{\pi}{\sqrt{67}}.
\end{equation}
In Fig.~\ref{fig:stablecuspyloop} we show 
the ${\vec p}$ and ${\vec q}$ curves for a stable loop with cusps similar
to~(\ref{eq:pqcuspy}) extracted from our simulations.

\section{Conclusions}
\label{conclusions}

We have shown that the fragmentation of a large complicated loop
yields stable quasi-planar, quasi-rectangular loops, that are
similar to degenerate kinky loops. If cosmic string loops are
the result of a large number of intercommutings -- even as few
as five or six intercommutings may be sufficient 
({\it e.g.} Fig.~\ref{fig:eigenvalue-average}) -- we 
expect that they too will be quasi-rectangular. Earlier work has
examined the stability of relatively simple loops (with a small
number of harmonics) and found stability over a region of
parameter space 
\cite{Copeland:1986kz,Chen:1987ve,Delaney:1989ne,Siemens:1994ir}.
It is unclear if cosmological evolution can directly produce 
these stable loops with few harmonics. Cosmological loops will 
certainly contain kinks and this feature is missing in the loops 
investigated in these earlier studies.  

In a cosmological setting, when a loop is produced from the infinite 
string network, it will fragment down to stable loops within two 
oscillation periods, and therefore only stable loops are relevant for 
cosmological signatures.  Indeed the
aim of the CA and SQSP studies was to obtain the gravitational power
emitted from realistic cosmic string loops. We can also estimate the
gravitational power based on the analytical results of \cite{Garfinkle1987}
for emission from degenerate, kinky loops, and obtain an estimate that is
within 13\% of the CA result. The gravitational power radiated
from degenerate kinky loops is analytically calculated to be 
$64\,{\rm ln}(2) G\mu^2 \approx 44 G\mu^2$ whereas CA numerically 
estimate $39 G\mu^2$ from their simulations, where $\mu$ is the string 
tension.

For cosmic strings, Hubble expansion, frictional damping and
radiation damping also come into play, though the effects are 
generally on very long time scales and depend on the environment 
in which the strings are placed. Hence we expect that when a 
large loop is produced from the network, it will fragment 
into stable loops within two oscillation periods and then, as these 
stable loops oscillate for many
oscillation periods, damping effects start to play a role and to
change the Nambu-Goto dynamics. So to characterize the effect of
damping on the cosmic string network loops, we can limit our study 
to the damping effects on the perturbed degenerate kinky loops
and the stable loops with cusps described above.

\begin{acknowledgments}
  We are grateful to Jose Blanco-Pillado, Ken Olum, Ben Shlaer
  and Alex Vilenkin for discussions and Isha Savani for help during the
  initial algorithm development. This work was supported by 
  the U.S. Department of Energy at Case Western Reserve University. 
  TV was also supported by grant number DE-FG02-90ER40542 at the 
  Institute for Advanced Study. 
  The numerical simulations were performed on the facilities provided 
  by the Case ITS High Performance Computing Cluster.
\end{acknowledgments}

\appendix

\section{Loop Intercommutation}
\label{app:loop-intercommutation}

\begin{figure}
  \includegraphics{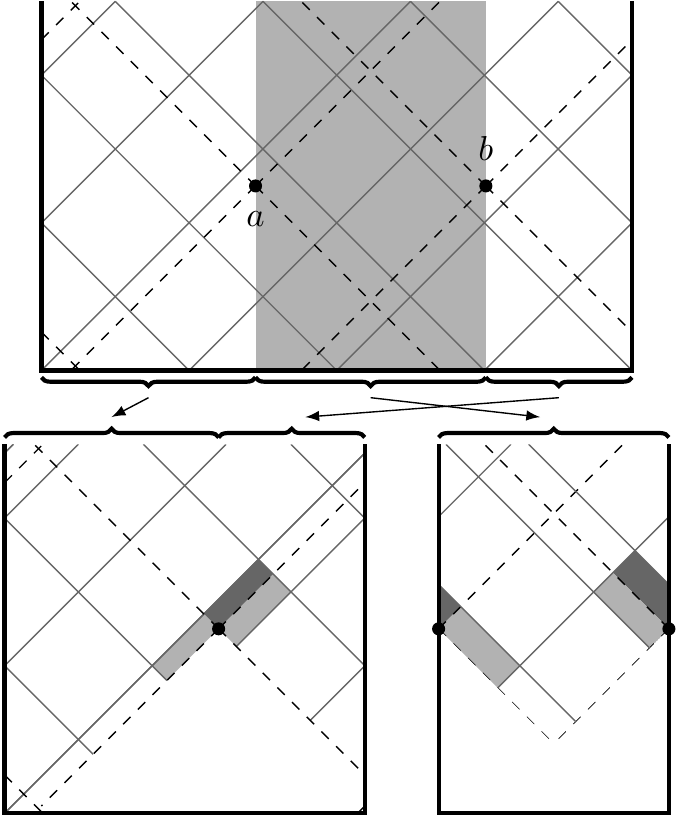}
  \caption{Geometric splitting of a loop.  Shown in this figure are the
    $\sigma_\pm$ grids for various loops.  The diagonal lines are lines of
    constant $\sigma_-$ (up and to the right) and constant $\sigma_+$ (up
    and to the left).  Time runs upward and $\sigma$ rightward.  The left
    and right edges of the grids are identified to describe a loop of
    string.  The initial loop is shown on the top with the intersection
    occurring at points $a$ and $b$.  The dashed lines show the constant
    values of $\sigma_\pm$ at which the intersection points occur and so at
    which kinks are formed.  The daughter loops are formed by cutting the
    grid and identifying the new edges.  In the figure the gray shaded
    region is cut from the grid and becomes one loop whereas the remaining
    unshaded pieces are connected.  The new loops are shown below the
    initial loop.  The past light cone of the intersection point has been
    removed since it no longer contains useful information.  The two
    intersecting diamonds are replaced by the three shaded diamonds in the
    figure.}
  \label{fig:loop-intersection}
\end{figure}

The numerical evolution in this work was performed using the ``diamond
method'' for string evolution which is discussed in~\cite{diamond-method}.
Full details of our implementation along with source code and documentation
is available online~\cite{copi-strings}.  Here we will discuss one aspect
of the evolution: loop intercommutation.

The diamond method gets its name from the $\sigma_\pm$ grid shown in
Fig.~\ref{fig:loop-intersection}.  The top of the figure shows the grid
structure for an initial loop.  Our initial loop segments are of uniform
length, $\Delta\sigma$, producing a regular grid.  This grid provides a
simple geometric picture of the loop.  With this grid structure, a starting
point, $\vec x(0,0)$, and values for $\vec p(\sigma_-)$ and $\vec
q(\sigma_+)$, taken to be piecewise constant functions, we can calculate
the coordinate of the string $\vec x(\sigma,t)$ during its evolution if
intersections do not occur.

In Fig.~\ref{fig:loop-intersection} we consider the case of an intersection
at points $a$ and $b$, that is, $\vec x(\sigma_a,\tint)=\vec x(\sigma_b,\tint)$.
In this picture the intercommutation and production of two loops is
geometric.  The shaded area is excised from the parent loop to become one
daughter loop (bottom right grid) and the remaining portions of the parent
loop are reconnected (bottom left grid).  The past light cone of the
intersection point no longer contains meaningful information and has been
removed from the new grids.  When an intersection occurs the grid structure
is modified; an originally uniform grid becomes non-uniform, but otherwise
contains the same information.  In the bottom left grid the light gray
shaded diamonds are the remnants of the original diamonds that
intersected.  The dark gray shaded diamond is a new one created by the
reconnection of the grid after the middle portion has been removed.  These
two grids now describe independent loops which may be evolved in the usual
way for all times $t>\tint$.

\section{Loop Rest Frame Transformation}
\label{app:transformation}

The time coordinate of a loop, $t$, was chosen to be the proper time and
the same as the background spacetime, $x^0=t$; see
Sec.~\ref{sec:procedure}.  To make $\vec p$ and $\vec q$ easier to boost we
construct four-vectors from them with the time components give by $p^0=-1$
and $q^0=1$. At a fixed time, $t$, we may then write in the initial loop's
rest frame
\begin{equation}
  p^\mu = \frac{\partial a^\mu}{\partial \sigma_-} = 
  \begin{pmatrix} -1 \\ \vec p \end{pmatrix},
\end{equation}
where $|\vec p|=1$.

The center of mass velocity of a loop is given by
\begin{eqnarray}
  \vcm & = & \frac12\left( \int \vec q\,d\sigma_+ - \int\vec
  p\,d\sigma_-\right) \\
  & = & \int \vec q\,d\sigma_+ = - \int\vec  p\,d\sigma_-. \nonumber
\end{eqnarray}
With $\gamma=1/\sqrt{1-\vcm^2}$ boosting to the rest frame of this loop we
find
\begin{equation}
  \label{eq:p-restframe}
  P^\mu(\sigma_-) 
   =  \begin{pmatrix} -\gamma (1+\vcm\cdot\vec p) \\ \vec p + (\gamma
    -1)(\vcm\cdot\vec p)\vcm/\vcm^2 + \gamma\vcm
  \end{pmatrix}.
\end{equation}

Notice that the time component of this four-vector is no longer $-1$.  To
correct this we apply a gauge transformation to the coordinate $\sigma_-$.  Let
\begin{equation}
  \tilde \sigma_- \equiv \gamma (\sigma_- + \vcm\cdot\vec a),
\end{equation}
so that
\begin{equation}
  \frac{\partial\tilde \sigma_-}{\partial \sigma_-} = \gamma (1 +
  \vcm\cdot\vec p).
\end{equation}
An integration gives $\tilde \sigma_- \in [0,1/\gamma ]$ and a coordinate 
rescaling can be used to bring the interval back to $[0,1]$.
With this we now have
\begin{eqnarray}
  P^0(\tilde \sigma_-) & = & \frac{\partial\tilde
    a^0}{\partial\tilde \sigma_-}
   = \left(\frac{\partial\tilde a^0}{\partial
    \sigma_-}\right)\left(\frac{\partial\tilde \sigma_-}{\partial
    \sigma_-}\right)^{-1}  \nonumber \\
   & = & \frac{P^0}{\gamma(1+\vcm\cdot\vec p)} = -1,
\end{eqnarray}
where we have used~(\ref{eq:p-restframe}) for $P^0$. Applying the
same transformation to the spatial piece of~(\ref{eq:p-restframe}) we find
\begin{equation}
  \vec{P}(\tilde\sigma_-) = \frac{\vec p+(\gamma-1)(\vcm\cdot\vec
    p)\vcm/\vcm^2 +\gamma\vcm}{\gamma(1+\vcm\cdot\vec p)}.
\end{equation}
Again this looks messy but we can verify that $|\vec{P}|=1$, as
required.  We can further verify that
\begin{equation}
  \int P(\tilde\sigma_-)\,d\tilde\sigma_- = \int
  P(\tilde\sigma_-)\left(\frac{\partial\tilde\sigma_-}{\partial\sigma_-}\right)
  d\sigma_- = 0.
\end{equation}

For $\vec q$ we proceed in the same way. In this case the gauge
transformation is $\tilde \sigma_+=\gamma(\sigma_+-\vcm\cdot\vec b)$ and we find
\begin{equation}
  \vec{Q}(\tilde\sigma_+) = \frac{\vec q+(\gamma-1)(\vcm\cdot\vec
    q)\vcm/\vcm^2 -\gamma\vcm}{\gamma(1-\vcm\cdot\vec q)}.
\end{equation}
Once again we can show that $|\vec{Q}|=1$.

The above equations are not in the best form for numerical evaluation.  If
$|\vcm|\approx 1$ then $\gamma\gg 1$ so large number will be subtracted
from each other in the numerator.  To correct this it is better to write
the equations in terms of $\gamma^{-1}=\sqrt{1-\vcm^2}$ since
$0\le\gamma^{-1}\le 1$. This provides the alternative forms
\begin{eqnarray}
  \vec{P} & = & \frac{\gamma^{-1}\vec p +
    (1-\gamma^{-1})(\vcm\cdot\vec p)\vcm/\vcm^2 + \vcm}{1+\vcm\cdot\vec p},
  \nonumber \\
  \vec{Q} & = & \frac{\gamma^{-1}\vec q + (1-\gamma^{-1})(\vcm\cdot\vec
    q)\vcm/\vcm^2 - \vcm}{1-\vcm\cdot\vec q}.
\end{eqnarray}

\section{Explicit Solutions of the String Equations and Constraints}
\label{newscheme}

Here we describe an explicit analytical solution of the Nambu-Goto
equations of motion, Eq.~(\ref{eqn:nambugoto}), and the string constraints,
Eqs.~(\ref{eqn:constraints}), (\ref{eqn:closure}).  By choosing the
decomposition in terms of left- and right- movers we can solve the
Nambu-Goto equations as in Eq.~(\ref{eqn:nambugoto}).  To solve the
constraint in Eq.~(\ref{eqn:constraints}) we start with
\begin{equation}
{\vec p}(\sigma_-) = (\sin\theta \cos\phi, \sin\theta \sin\phi, \cos\theta)
\end{equation}
where $\theta$ and $\phi$ are functions of $\sigma_-$.
We similarly choose ${\vec q}(\sigma_+)$ with independent
angular functions.

We require that ${\vec p}$ be periodic under $\sigma_- \to \sigma_- +
1$. (Throughout this section we consider a loop of length $1$.)  Therefore
we write
\begin{equation}
\theta (\sigma_-) = 2 \pi j_- \sigma_- + 
             \sum_{m=0}^\infty \theta_m \cos (2 \pi m \sigma_- + \alpha_m)
\end{equation}
where $j_-$ is an integer, and $\theta_m$ and $\alpha_m$ are arbitrary 
constants that can be chosen randomly to generate random loops. Similarly,
\begin{equation}
\phi (\sigma_-) = 2\pi k_- \sigma_- + 
             \sum_{m=0}^\infty \phi_m \cos ( 2 \pi m \sigma_- + \beta_m)
\end{equation}
where $k_-$ is an integer, and $\phi_m$ and $\beta_m$ are arbitrary 
constants. 

Next we come to the closure condition in Eq.~(\ref{eqn:closure}).  The
integral of ${\vec p}$ will, in general, not vanish. To correct this, 
define
\begin{equation}
{\vec v}_- \equiv - \int {\vec p}(\sigma_-) d\sigma_-
\end{equation}
and boost ${\vec p}$ to velocity ${\vec v}_-$ followed by a gauge
transformation as described in Appendix~\ref{app:transformation}.
This gives us the final solution,
\begin{equation}
\vec{P}(\tilde\sigma_-) =
 \frac{\vec p + (\gamma_--1)({\vec v}_- \cdot\vec p){\vec v}_-/{\vec v}_-^2
                           +\gamma_- {\vec v}_-}
       {\gamma_- (1+{\vec v}_-\cdot\vec p)}
\end{equation}
where $\gamma_-= (1-{\vec v}_-^2)^{-1/2}$ and
\begin{equation}
d\tilde\sigma_- = \gamma_- (1+{\vec v}_- \cdot {\vec p}) d\sigma_-
\end{equation}
As in Appendix~\ref{app:transformation}, an integration shows that
$\tilde \sigma_- \in [0,1/\gamma ]$ and a coordinate rescaling can
be used to bring the interval back to $[0,1]$.

It is straightforward to check directly that $|{\vec P}| =1$ and also 
\begin{equation}
\int \vec{P} (\tilde \sigma_-) d{\tilde\sigma}_- = 0
\end{equation}
In a similar way we construct ${\vec Q} (\tilde \sigma_+)$.

This scheme has the advantage that it solves the constraints
exactly and explicitly though it still requires integrating
${\vec p}(\sigma_-)$ to find ${\vec v}_-$ and ${\tilde \sigma}_-$.
We have not employed this scheme in the work reported here but
have remained with the SP algorithm to facilitate comparison.

\bibstyle{aps}
\bibliography{paper}

\end{document}